\documentstyle[emulateapj,pstricks]{article}
\def\ap3m{\mbox{AP$^3$M}}

\def\delth{\mbox{$\bar \delta_{\rm TH}$}}

\def\ddm{\mbox{$\delta_{\rm dm}$}}

\def\kms{\mbox{km/s}}

\def\kpch{\mbox{$h^{-1}$kpc}}

\def\lcdmtt{\mbox{\char'3CDM$_{30}~$}}
\def\LCDM{\mbox{\char'3CDM}}
\def\lcdmxx{\mbox{\char'3CDM$_{60}$~}}
\def\lcdm30art{\mbox{\char'3CDM$_{30}^{\rm ART}$~}}
\def\mhalo{\mbox{$M_{\rm min}$}}

\def\mpch{\mbox{$h^{-1}$Mpc}}

\def\msunh{\mbox{$h^{-1}$M$_\odot$}}

\def\nstep{\mbox{$N_{\rm steps}$}}
\def\ome{\mbox{$\Omega_0$}}
\def\omeb{\mbox{$\Omega_b$}}
\def\sig8{\mbox{$\sigma_8$}}
\def\tCDM{\mbox{$\tau$CDM}}

\def\xidm{\mbox{$\xi_{dm}$}}
\def\xihh{\mbox{$\xi_{hh}$}}
\def\xigg{\mbox{$\xi_{gg}$}}

\def\mathnew{\mathsurround=0pt}
\def\ref{\par\noindent\hangindent=2pc \hangafter=1 }
\def\simov#1#2{\lower .5pt\vbox{\baselineskip0pt
    \lineskip-.5pt\ialign{$\mathnew#1\hfil##\hfil$\crcr#2\crcr\sim\crcr}}}

\def\'#1{\ifx#1i{\accent"13\i}\else{\accent"13#1}\fi}
\def\eg{e.g.,}
\def\et{et~al.}

\begin{document}
%\submitted{submitted to the Astrophysical Journal}
\slugcomment{{\em submitted to the Astrophysical Journal}}
\lefthead{EVOLUTION OF BIAS}
\righthead{COLIN ET AL.}

\title{Evolution of bias in different cosmological models}

\author{Pedro Col\'in}
\affil{Instituto de Astronom\'ia, Universidad Nacional Aut\'onoma
de M\'exico, C.P. 04510, M\'exico, D.F., M\'exico}
\vspace{1mm}
\author{Anatoly A. Klypin, and Andrey V. Kravtsov}
\affil{Astronomy Department, New Mexico State University, Box 30001, Dept.
4500, Las Cruces, NM 88003-0001}
\vspace{1mm}
\author{Alexei M. Khokhlov}\vspace{2mm}

\affil{Laboratory for Computational Physics and Fluid Dynamics, Code
  6404, 
Naval Research Laboratory, Washington, DC 20375}

\begin{abstract}
  
  We study the evolution of the halo-halo correlation function and bias in
  four cosmological models (\LCDM, OCDM, \tCDM, and SCDM) using very
  high-resolution $N$-body simulations with dynamical range of $\sim
  10,000-32,000$ (force resolution of $\approx 2-4\kpch$ and particle
  mass of $\approx 10^9h^{-1} {\rm M_{\odot}}$).  The high force and
  mass resolution allows dark matter (DM) halos to survive in the tidal
  fields of high-density  regions and thus prevents the
  ambiguities related with the ``overmerging problem.'' This allows us
  to estimate for the first time the evolution of the correlation function
  and bias at small (down to $\sim 100\kpch$) scales.
  
  We find that at all epochs the 2-point correlation function of
  galaxy-size halos $\xihh$ is well approximated by a power-law with
  slope $\approx 1.6-1.8$.  The difference between the shape of $\xihh$
  and the shape of the correlation function of matter results in the
  {\em scale-dependent bias} at scales $\lesssim 7\mpch$, which we find
  to be a generic prediction of the hierarchical models, independent of
  the epoch and of the model details. The bias evolves rapidly from a
  high value of $\sim 2-5$ at $z\sim 3-7$ to the {\em anti-bias} of
  $b\sim 0.5-1$ at small $\lesssim 5\mpch$ scales at $z=0$.  Another
  generic prediction is that the {\em comoving} amplitude of the
  correlation function for halos above a certain mass evolves
  non-monotonically: it decreases from an initially high value at
  $z\sim 3-7$, and very slowly increases at $z\lesssim 1$.  We find
  that our results agree well with existing clustering data at
  different redshifts, indicating the general success of the
  hierarchical models of structure formation in which galaxies form
  inside the host DM halos.  Particularly, we find an excellent
  agreement in both slope and the amplitude between $\xihh(z=0)$ in our
  \lcdmxx simulation and the galaxy correlation function measured using
  the APM galaxy survey. At high redshifts, the observed clustering of
  the Lyman-break galaxies is also well reproduced by the models.  We
  find good agreement at $z\gtrsim 2$ between our results and
  predictions of the analytical models of bias evolution. This
  indicates that we have a solid understanding of the nature of the
  bias and of the processes that drive its evolution at these epochs.
  We argue, however, that at lower redshifts the evolution of the bias
  is driven by dynamical processes inside the nonlinear high-density
  regions such as galaxy clusters and groups.  These processes do not
  depend on cosmology and tend to erase the differences in clustering
  properties of halos that exist between cosmological models at high
  $z$.
\end{abstract}

%=====================

\section{Introduction}

%=====================

It is widely believed that the distribution of galaxies is different from
the overall distribution of dark matter (DM). This difference, {\em the
  bias}, is crucial for comparisons between observations and
predictions of cosmological models.  Observations provide information
about the distribution of {\em objects} such as galaxies and galaxy
clusters. The models, however, most readily predict the evolution of the dark
matter distribution, which cannot be observed directly. The models,
therefore, should be able to predict the distribution of {\em objects} or,
conversely, predict how this distribution is different from that of
the dark matter (i.e., the bias). The notion of bias was introduced by
\cite{kaiser84} to explain the large difference in clustering strength
between galaxies and Abell clusters.  Later \cite{kaiser86} and
\cite{bardeen86} applied this argument to galaxies themselves.  Davis
et al. (1985) showed that the CDM model disagreed with observations, if
galaxies were distributed like dark matter.  However, if a biased
galaxy formation scenario was assumed, the ``galaxy'' correlations
substantially exceeded the correlations of mass at all scales and
agreed with observations for certain values of bias.  This work was
followed by other studies which tried to account for the phenomenon of
galaxy bias (\eg\ \cite{rees85}; \cite{ss85}; \cite{silk85};
\cite{ds86}).

The bias can be defined and understood differently. In this
paper we will use the conventional statistical definition of the bias as
the ratio of the correlation functions of objects and dark
matter:
\begin{equation}
b^2(M,r,z) \equiv \frac{\xigg(M,r,z)}{\xidm(r,z)}.
\end{equation}
Here, $b^2$ is the square of the {\em bias function}, $\xigg(M,r,z)$
and $\xidm(r,z)$ are the 2-point spatial correlation functions of
objects and dark matter, respectively. Dependencies in the above
equation indicate that in general the bias may depend on the epoch $z$,
scale $r$, and properties of the objects such as their mass $M$. It is
also expected that the functional form $b^2(M,r,z)$ depends on the
cosmological model.  The bias in the above definition is most closely
related to the observations because the {\em observed} $\xigg(M,r,z)$
can be compared to the $\xidm(r,z)$ {\em predicted} within a framework
of a given cosmological model. This gives an estimate of {\em the
  amount of bias needed} for a cosmological model to agree with
observations.  While it is clear that this approach cannot be used to
test a model, it provides some insights into the nature of bias and its
evolution. For example, Steidel et al. (1998) used the observed
clustering strength of high-redshift ($z\approx 3$) Lyman-break
galaxies to derive the implied value of bias in different cosmological
models. This value was found to be quite large in all models ($b\sim
3-6$). On the other hand, a number of theoretical studies (e.g.,
Klypin, Primack \& Holtzman 1996; Cole et al. 1997; Jenkins et al.
1998; and references therein) estimated \xidm\ in different
cosmological models at $z=0$ and made comparisons with accurate
measurements of \xigg\ from local galaxy surveys.  These studies
indicate that significant {\em anti-bias} ($b < 1$) is required for the
open and flat low-\ome\ models at scales $\lesssim 3-8\mpch$, while
$\Omega_0=1.0$ cluster-normalized models indicate positive bias ($b >
1$).  Comparison of the low- and high-$z$ results implies that,
regardless of cosmological model, the bias has decreased significantly
from the early epochs to the present.

The distribution of the dark matter and $\xidm(r,z)$ cannot be observed
directly. Therefore, a test of a cosmological model is possible only if
the model can predict the $\xigg$ of observed objects. Unfortunately, this
is not an easy and straightforward task. First of all, we should fully
understand what are the observed objects and where/how they form. The
standard lore is that observed galaxies form dissipatively inside dark
matter halos. The properties of galaxies will then depend on the mass
of the parent halo, its spin, details of dissipative processes and mass
accretion history, and other factors (e.g., Mo, Mao \& White 1998).
Therefore, in order to predict the type of galaxy and its properties,
the relevant processes must be included in the model. However, it seems
likely that in {\em every} sufficiently massive ($M\gtrsim
10^{11}h^{-1} {\rm M_{\odot}}$) {\em gravitationally bound} halo
baryons will cool, form stars, and produce
an object ressembling a galaxy (e.g., Kauffman, Nusser \& Steinmetz 1997; Roukema et
al. 1997; Yepes et al.  1997; \cite{sp97}). The galaxy population {\em
  as a whole} can be then viewed as a population of {\em galaxy-size}
dark matter halos. The clustering of the latter can be studied without
inclusion of complicated physics; it has been modelled using both direct
numerical simulations and analytical methods.

Typical ingredients of the analytical models (e.g., Matarrese et al.
1997; Moscardini et al. 1998; \cite{Mann97}) are the extended
Press-Schechter formalism (\cite{bower91}; \cite{bond91}) used to
follow mass evolution of halos, analytical approximations to the
non-linear clustering evolution of the dark matter (e.g., Hamilton et
al. 1991; Peacock \& Dodds 1994,1996; Jain, Mo \& White 1995; Smith et
al. 1998), and a model for bias evolution (e.g., Mo \& White 1996;
Matarrese et al.  1997). The evolution of bias in such models can be
calculated quickly which makes extensive parameter-space
studies possible. The main disadvantages are large uncertainties (especially at
small, $r\lesssim 5h^{-1} {\rm Mpc}$, scales) introduced by bias
prescriptions and limited applicability of the non-linear clustering
approximations.

Direct numerical simulations should, ideally, predict halo-halo
clustering without any additional assumptions and uncertainties.
However, until very recently the predictions of numerical simulations
were also quite uncertain.  The main reason for the uncertainty was
that dissipationless $N$-body simulations had been consistently failing
to produce galaxy-size dark matter halos in dense environments typical
for galaxy groups and clusters.  Recently, it was shown that this
effect, known as ``the overmerging problem'' (\eg\ \cite{frenk88};
Summers, Davis, \& Evrard 1995), is due mainly to the insufficient
force and mass resolution of such simulations (Moore, Katz \& Lake
1996; \cite{kgkk97}, hereafter KGKK; \cite{ghigna98}).  The lack of
sufficient resolution leads to artificial disruption of halos in
clusters. This, in turn, leads to a strong artificial anti-bias
(especially at small scales $\lesssim 3h^{-1} {\rm Mpc}$, but larger
scales are also affected).  There are several ways to deal with this
problem. One possible way is to break up massive structureless halos
into subhalos using some kind of observationally motivated prescription
(\eg\ \cite{nkp97}; Klypin, Nolthenius, \& Primack 1997).  These
subhalos can then be included into halo catalogs used to compute the
correlation function and other halo statistics.  Another common
approach is to overcome overmerging by weighting the massive halos
according to their mass and compute thus a {\em weighted} correlation
function (\eg\, \cite{bagla97}). Both of these approaches are useful.
Nevertheless, it is not clear whether the unavoidable heuristic
assumptions take the processes driving the small-scale bias evolution
correctly into account.  The weighting technique, for example, ignores
the real physical effects in groups and clusters such as tidal
stripping and dynamical friction.
We will argue below that these effects are
likely to be driving the small-scale bias evolution at low ($z\lesssim
1$) redshifts. Hydrodynamic simulations that include gas cooling are
affected by overmerging to a significantly lesser degree (e.g., Summers
et al. 1995; \cite{khw98}). The cooling creates compact dense objects
inside halos which can survive in clusters. These simulations,
therefore, can be used to study the halo clustering directly.
Unfortunately, the computational cost required to simulate a large
volume with sufficiently high mass resolution is prohibitively high.
Moreover, such simulations usually oversimplify the gas dynamic by
including only cooling mechanism. This leads to ``overcooling'':
without a heating process to regulate it, the cooling produces very
compact and dense baryonic blobs in the halo centers. These blobs do
survive successfully in clusters, but they also suffer much less from
the tidal stripping of material as compared to a realistic galaxy with
a more extended distribution of baryons. The mass of the objects may
thus be higher than it should be which may lead to excessive dynamical
friction and thus incorrect dynamics of halos.

The very high resolution $N$-body simulations are thus a viable
alternative, if the required resolution can be reached at an affordable
computational cost.  Analytical arguments and numerical experiments
(Moore et al. 1996; KGKK) indicate that the required force and mass
resolution are $\approx 1-2h^{-1} {\rm kpc}$ and $\lesssim 10^9h^{-1}
{\rm M_{\odot}}$, respectively. This force resolution is sufficiently
high for most halos to survive in high-density regions.  The mass
resolution is determined by the requirement that a galaxy-size halo
should have $\gtrsim 100$ particles to avoid relaxation effects and to
assure a robust identification by a halo finding algorithm. The
numerical simulations that reach such resolution (Ghigna et al. 1998;
KGKK) show that most halos do survive even in the richest clusters. The
dynamic range required to reach the needed resolution in a
statistically large volume, $\sim 50-100h^{-1} {\rm Mpc}$, is quite
high: $2-5\times 10^4$. Nevertheless, the advances in computer hardware
and in the numerical algorithms make it now possible to carry out such
simulations at an affordable computational cost. In this paper we use
such high dynamic range simulations to calculate the 2-point
correlation function of halos and corresponding bias using {\em all}
halos in the simulated volume: i.e., both isolated dark matter halos
and satellites of massive halos, and halos inside group- and
cluster-size systems.  The simulations of different cosmological models
and box sizes were made using the Adaptive Refinement Tree (ART;
Kravtsov, Klypin \& Khokhlov 1997) and the AP$^3$M (Couchman 1991)
$N$-body codes. The absence of the overmerging\footnote{The extent to
  which the very central regions ($r\lesssim 200h^{-1} {\rm kpc}$) of
  clusters are affected by the overmerging, even with resolution this
  high, is still a matter of debate. Ghigna et al. (1998), for example,
  argue that these regions are still completely overmerged. They find,
  however, that halos survive at smaller radii ($\approx 50h^{-1}$ kpc)
  in their $5$ kpc resolution RUN1, as compared to the $10$ kpc RUN2.
  The resolution of the simulations presented in this paper is
  comparable to the RUN1. We do find halos within central $\approx
  100h^{-1} {\rm kpc}$ (see \S~4.3 and Figs. 1-3). Although some halos
  do survive at $r\lesssim 100h^{-1} {\rm kpc}$ from cluster center,
  our tests show that these regions may be affected by the overmerging.
  The larger scales $\gtrsim 100h^{-1} {\rm kpc}$, are not affected} in
these simulations means that the additional steps such as breaking-up
of clusters, or mass-weighting (see above) are not necessary.
Therefore, the correlation function of halos is measured {\em directly}
down to unprecedentedly small scales ($\approx 150h^{-1} {\rm kpc}$)
without the usual uncertainties associated with these steps. For the
first time this opens the possibility to study the effects of dynamical
processes such as tidal destruction and dynamical friction on the
evolution of small-scale bias.  The results on the evolution of bias
and on the halo correlation function can be used as a basis for
comparisons and interpretations of the existing and upcoming
observations, as well as a check and/or input for the analytic models
of clustering evolution.

The paper is organized as follows. In \S~2 we briefly review the
definitions of bias and current analytical models of its evolution.
The cosmological models studied in this paper are described in \S~3.
The details of the numerical simulations and discussion of the
construction of halo catalogs and halo survival in the high-density
regions is given in \S~4.  In \S~5 we present our results on the
evolution of halo clustering and bias in different cosmological models.
We also present a comparison of our $z=0$ results with the galaxy
correlation function measured using the APM galaxy survey.  A
discussion of the main results is presented in \S~6. We summarize our
main results and conclusions in \S~7.

%===========================

\section{The notion of bias}

%===========================

The notion of bias is more complicated than eq. (1) might suggest.
Formally, the bias is defined as a function relating fluctuations in
the dark matter density, $\delta_{dm}$, to the fluctuations in the
number density of objects, $\delta_n$. In general, this relationship
can be complicated and may depend on a large number of factors:
scale, mass and/or type of objects, time, etc.  The large number of the
dependencies or discreteness of the object distribution can result in
the {\em stochasticity} of the bias (Dekel \& Lahav 1998): a scatter in
the relationship between $\delta_n$ and $\delta_{dm}$. Finally, the
bias can be {\em non-local}, if probability to form an object at a
given point is not fully determined by local factors. For example, in
addition to the local density, temperature, etc., the probability to
form a galaxy may depend on environment. Analytical results of Catelan
et al. (1998b) indicate that the bias is expected to be non-local even in
the linear regime.

The lack of general understanding of these dependencies of the bias usually
results in the use of the simplest assumptions.  The most common
approach is to assume that bias is {\em local}, depends only on the
local matter density, and is {\em linear}: $\delta_n=b\delta_{dm}$. An
obvious consequence of the latter assumption is that bias is
scale-independent. In this case, the correlation function and power
spectrum of halos and DM are simply related by a constant scaling
factor. Although the linearity of bias would greatly simplify the
theoretical interpretation of the clustering data, it is likely that
bias depends on a variety of processes that may lead to
nonlinearity. These processes include merging, tidal disruption (e.g.,
Dubinski 1998), suppression of galaxy formation in small halos due to
supernovae feedback (Dekel \& Silk 1986; Yepes et al. 1997), etc.
The observations, in fact, indicate that at small scales galactic bias {\em is}
nonlinear. The correlation functions of different types of galaxies
differ in amplitude and shape suggesting that at least some of the
galaxies are nonlinearly biased.

During the last years, a significant progress has been made in the
analytical modelling of the bias and its evolution (e.g., Coles 1993;
Fry 1996; Mo \& White 1996, hereafter MW; Matarrese et al. 1997; Mann
et al. 1997; Catelan et al. 1998a,b). Although the current analytical
models may have some drawbacks and limitations, they provide an insight
and interpretation for numerical simulations.  At this point it is also
important to check how well the prediction of the analytical models
agree with the results of simulations (see, e.g., MW; Mo, Jing \& White
1996; Jing 1998). We will therefore review briefly some results of the
analytical models concerning evolution of the bias.

In a seminal paper, Kaiser (1984) showed that if the observed systems in the
universe (such as galaxies and galaxy clusters) form in the
{\em peaks} of the density field, their distribution is biased.  This
is an example of {\em statistical} bias: the bias determined simply by
the fact that objects do not sample the distribution of matter but that
of the peaks, the latter having a statistically different distribution
from the former. Thus, the bias is introduced from the start by the
way the objects form. However, the bias changes subsequently as
the distribution evolves driven by the gravitation. In hierarchical
models this evolution results in growth of objects via multiple
mergers. However, if the merger rate is low, the evolution of bias 
can be modeled by the {\em object-conserving model}. In this model the
objects form with an initial statistical bias and after that are
dragged without merging by a gravitational pull from the surrounding
density fluctuations (Dekel \& Rees 1987; Nusser \& Davis 1994; Fry 1996).

MW used the extended Press-Schechter formalism to derive an expression
for the bias in Lagrangian coordinates\footnote{More sophisticated
  analytical treatment of bias in Lagrangian coordinates can be found
  in Catelan et al. (1998b) and Mann et al. (1997).  } (comoving radius
$R$ of the region from which halos form or halo mass $M$). At linear
scales, this expression\footnote{MW note that equation (2) should be
  valid even when $\ddm \gtrsim 1$ or $\xi_{dm}(r)\sim 1$, as long as
  the scale $r$ is larger than the Lagrangian radius $R$.  They find
  that at these scales the correlation function of halos and the bias in their
  numerical simulations are well described by equation (2) (see,
  however, Catelan et al. 1998a, and Jing 1998).}  for the bias at
redshift $z$ for DM halos of mass $M$ formed at redshift $z_f$ is
(Matarrese et al.  1997):
\begin{equation}
b(M,z|z_f) = 1 + \frac{\nu^2 -1}{\delta_f}.
\end{equation}
Here $\delta_f = \delta_c D_+(z) /D_+(z_f)$, $D_+(z)$ is the growing
mode of linear perturbations normalized to unity at $z=0$,
$\nu=\delta_f/\sigma(M,z)$, $\delta_c$ is the critical overdensity for
spherical collapse at $z=0$, and $\sigma(M,z)$ is the rms linear mass
fluctuation on the scale $M$ of halos linearly extrapolated to redshift
$z$.  This expression is similar to that obtained in earlier studies
from a peak-background split argument (e.g., Cole \& Kaiser 1989).  The
standard interpretation of the Press-Schechter description of the
hierarchical evolution is that at any epoch $z$ all halos merge
immediately to form more massive halos. Thus, if observed objects are
identified with host halos at any epoch, then $z=z_f$ in eq. (2).
Matarrese et al. (1997) call this {\em the merging model}. The
object-conserving and merging models are two extremes pictures of
clustering evolution, although they may be applicable to the evolution
of galaxy clustering at certain epochs. At some epochs the halos
may neither survive nor merge instantly. Moreover, other processes such
as halo dynamics in galaxy clusters and groups may become important
when clustering reaches highly nonlinear stages.  The wealth of
potentially important processes may make a one-to-one identification
between galaxies\footnote{The situation with evolution of clusters of
  galaxies is, of course, much simpler.} and DM halos very difficult
(see, for example, discussion in Moscardini et al. 1998). 

Eq. (2) gives an estimate of the bias for objects of a single
mass. The bias of a sample of objects (``effective'' bias\footnote{We
  follow here the notation and terminology of Matarrese et al.
  (1997).}) with a range of masses $M>M_{min}$ should be calculated as
a weighted {\em average} over the mass distribution of objects
$n(M,z)$:
\begin{equation}
b_{eff}(z) =n(z)^{-1}\int\limits_{M>M_{min}} b(M,z|z_f)n(M,z)
d\ln M,
\end{equation}
where $n(M,z)$ is given by the Press-Schechter (1974) distribution and $n(z)$
is the mean number density of objects with masses $M>M_{min}$ at
redshift $z$.  Moscardini et al. (1998) give a useful fitting formula
for $b_{eff}(z)$:
\begin{equation}
b_{eff}(z) = 1-1/\delta_c + [b_{eff}(0)-
1 + 1/\delta_c]/D_+(z)^{\beta},
\end{equation}
and provide best-fit parameters $b_{eff}(0)$ and $\beta$ for a variety
of different cosmological models and values of $M_{min}$.  The generic
feature of the evolution of the effective bias described by eq.  (4) is its
rapid decrease with decreasing redshift and increase with increasing
$M_{min}$. For galaxy-size halos, $b_{eff}(0)\sim 0.5-1.0$ and
$\beta\sim 2.0-1.7$ for $M_{min}\sim 10^{9}-10^{12}h^{-1} {\rm
  M_{\odot}}$. This gives $b_{eff}\sim 2-4$ at $z=3$ and $b_{eff}\sim
0.5-1.0$ at the present epoch. We will use eq. (4) in \S~6 to interpret
and compare our results with predictions of the analytical models
discussed above.

\begin{deluxetable}{lcccccl}
%\tablewidth{58pc}
\tablecolumns{7}
\tablecaption{Cosmological Models}
\tablehead{ \colhead{Model} & \colhead{\ome} & 
\colhead{$\Omega_{\Lambda,0}$} & \colhead{$h$} & 
\colhead{$t_0$} & \colhead{\sig8} & \colhead{Approximation}\\ 
& & & &  (Gyr)& & }
\startdata
SCDM\phm{100} & 1.0 & 0.0 & 0.50 & 13.1 & 1.1 & Efstathiou et al.\nl 
OCDM\phm{100} & 0.3 & 0.0 & 0.65 & 12.2 & 0.9 & BBKS+Sugiyama\nl
\char'3CDM$_1$\phm{100} & 0.3 & 0.7 & 0.70 & 13.4 & 1.17 & BBKS+Sugiyama\nl
\char'3CDM$_2$\phm{100} & 0.3 & 0.7 & 0.70 & 13.4 & 1.0 & Klypin \& Holtzman \nl
\tCDM\phm{100} & 1.0 & 0.0 & 0.5 & 13.1 & 1.0 & Efstathiou et al.\nl
\enddata
\end{deluxetable}

%============================

\section{Cosmological models}

%============================

We have chosen to study the evolution of bias in four representative
variants of the CDM models (see details in Table 1): 1) the standard
{\sl COBE}-normalized CDM model (SCDM); 2) a variant of the $\ome =
1.0$ CDM model (\tCDM) with a different shape of the power spectrum
(the shape parameter $\Gamma = \ome h = 0.2$); 3) a flat low-density
model with $\ome = 1-\Omega_{\Lambda}=0.3$ (\LCDM); 4) an open model
with $\ome = 0.3$ (OCDM). The observations of the galaxy clustering indicate
that the power spectrum of the galaxy distribution has a shape different from
that of the standard CDM model: $\Gamma=\Omega_0 h\approx 0.2$ instead
of $\Gamma = 0.5$ of the SCDM (e.g., Maddox, Efstathiou, \& Sutherland
1996; Peacock \& Dodds 1994). This has motivated Jenkins et al. (1998)
to study the {\tCDM} model, in which $\tau$ indicates the fact that
lower values of $\Gamma$ may be obtained with a late decay of the
massive $\tau$-neutrino. The physics (additional to the physics of the
SCDM model) responsible for the change of the shape is, in fact,
irrelevant for study of structure formation. Although the model in that
respect is somewhat heuristic, it is interesting as an example of a
model with $\Omega_0=1$ and an approximately correct shape of the power
spectrum. The observations of the galaxy cluster evolution (Eke et al.
1998) and of the baryon fraction in clusters (\cite{evrard97}) strongly
indicate value of matter density $\ome \approx 0.3$, while various
observational measurements of the Hubble constant (e.g., Kim et al.
1997; Falco et al. 1997; Salaris \& Cassisi 1998) tend to converge on
the values of $h\approx 0.6-0.7$. Therefore, we have considered two
models (open and flat) with $\Omega_0=0.3$ and $h=0.65-0.7$. The age of
the Universe in all of our models is given in Table 1 and is in good
agreement with the ages of the oldest globular clusters (Chaboyer
1998).

We have used different approximations for the power spectrum of density
fluctuations for different considered models. For the OCDM and
\LCDM$_1$ models (see Table 1) we used the Bardeen et al. (1986; BBKS)
fit for the power spectrum $P(k)= A k T^2 (k)$ with corrections of \cite{sugiyama95}:
\begin{eqnarray}
T(k)&=&\frac{\ln(1+2.34q)}{2.34q}\times[1+3.89q+\nonumber \\
    & &(16.1q)^2 + (5.46q)^3+(6.71q)^4]^{-1/4}
\end{eqnarray}
and
\begin{equation}
q = k \left[\ome h^2 \exp[-\Omega_b -\sqrt{2h} (\Omega_b/\ome) \right]^{-1}
\end{equation}
where $\Omega_b = 0.0125 h^{-2}$ (\cite{walker91}\footnote{For the
  \LCDM$_2$ we used a slightly higher value of $\Omega_b = 0.015h^{-2}$.
  }). The power spectra
for both the SCDM and the \tCDM\ models were approximated by the fitting
formula of \cite{ebw92}
\begin{equation}
P(k)= \frac{Ak}{\left( 1+[ak + (bk)^{3/2} + (ck)^2]^{1.13}
  \right)^{2/1.13}}
\end{equation}
where $a=6.4/\Gamma$, $b=3.0/\Gamma$, $c=1.7/\Gamma$, and $A$ is the
normalization constant.  The shape parameter $\Gamma$ is $0.5$ and
$0.2$ for the SCDM and {\tCDM} models, respectively.  These two
analytic fits provide fairly good approximations to the power spectra
of these models in the limit $\omeb/\ome \ll 1$.

For the \LCDM$_2$ models we have used an approximation to the power
spectrum different from that of the \LCDM$_1$ model.  The
approximation
\begin{equation}
P(k)= \frac{Ak}{\left(1+a_1k^{1/2}+a_2k+a_3k^{3/2}+a_4k^2)
  \right)^{2a_5}},
\end{equation}
where $a_1=-1.5598$, $a_2=47.986$, $a_3=117.77$, $a_4=321.92$, and
$a_5=0.9303$, 
is given by Klypin \& Holtzman (1997) and was obtained by a direct fit
to the power spectrum estimated using a Boltzmann code. The accuracy of
this approximation is $\lesssim 2\%$ (see Klypin \& Holtzman 1997 for
details). A small shift in the normalization makes the approximations
for the $\LCDM_1$ and $\LCDM_2$ models very similar in the range of
wavenumbers probed in our simulations ($k\approx (0.1-10)h{\rm
  Mpc}^{-1}$).  This shift can be expressed using the value of \sig8
(amplitude of fluctuations on $8\mpch$ scale): $\sig8=1.17$ for the BBKS
approximation  versus $\sig8=1.0$ for the Holtzman
approximation. With these values of {\sig8} the differences in power
are negligible at large scales ($k\approx (0.1-0.3)h{\rm Mpc}^{-1}$)
and are within 10\% at the smaller scales.

Our SCDM model was normalized to the two-year {\sl COBE}-DMR data,
$\sig8 = 1.1$ (a somewhat higher value, $\sig8\approx 1.15-1.2$, is obtained
from the four-year COBE-DMR data, see e.g., Bunn \& White 1997). This
normalization is inconsistent with normalization, $\sigma_8\approx
0.5$, deduced from the observed cluster abundances (e.g.,
\cite{ecf96}), which reflects the well-known failure of this
model to account for both {\sl COBE} and cluster data.  Our
normalization of the OCDM is higher than that implied by the four-year
{\sl COBE}-DMR data ($\sigma_8\approx 0.5$), but is
consistent with the normalization, $\sigma_8\approx 0.9$, implied by
the cluster abundances. The normalization of our $\Lambda$CDM models,
on the other hand, is in good agreement with both cluster and the
4-year {\sl COBE} data. As explained above, the small difference in
normalization between the $\LCDM_1$ and $\LCDM_2$ was introduced to
minimize differences between the two power spectrum approximations used
in these models. Normalization of the {\tCDM} model is inconsistent
with both the {\sl COBE} normalization ($\sigma_8\approx 0.45$) and
with cluster abundance normalization ($\sigma_8\approx 0.52$). This is
motivated by our intent to use {\tCDM} as a toy-model rather than a
reasonable approximation to the real universe. The normalization of the
{\tCDM} is similar to normalizations of the CDM and {\LCDM} models.
Therefore, comparison between results of these models allows us to
study effects of changing the shape of the power spectrum and value of
$\Omega_0$.

%===================================

\section{The numerical simulations}

%===================================

\begin{deluxetable}{lllccrcccc}
%\tablewidth{58pc}
\tablecolumns{9}
\tablecaption{Parameters of simulations}
\tablehead{\colhead{Code}\hfil & \colhead{Model}\hfil &  \colhead{Run}\hfil & \colhead{$z_{\rm init}$} 
 & \colhead{$m_{\rm particle}$} & \colhead{\nstep} & \colhead{Resolution} & Box & $N_{\rm part}$\\ 
 &  & && (\msunh)& & (kpc/h)& (Mpc/h) & &}
\startdata
\ap3m & SCDM\phm{100}& SCDM\phm{100} & 49\phm{100} & $3.5\times 10^9$  & 8000 & 3.0 & 30 &$128^3$\nl 
\ap3m & OCDM\phm{100}& OCDM\phm{100} & 109\phm{100} & $1.1\times 10^9$ & 7000 & 4.7 & 30 &$128^3$\nl
\ap3m &\char'3CDM$_1$\phm{100}& \lcdmtt\phm{100} & 64\phm{100} & $1.1\times 10^9$  & 4000 & 3.0 & 30 &$128^3$\nl
ART &\char'3CDM$_2$\phm{100}&\lcdmxx\phm{100} & 30\phm{100} & $1.1\times 10^9$ & 41300 & 1.8 & 60 &$256^3$\nl
ART &\char'3CDM$_2$\phm{100}&\lcdm30art\phm{100} & 45\phm{100} & $1.3\times 10^8$ & 13800 & 0.9& 30 &$256^3$ \nl
\ap3m & \tCDM\phm{100}& \tCDM\phm{100} & 50\phm{100} & $3.5 \times 10^9$ & 8000 & 3.0 & 30 &$128^3$\nl
\enddata
\end{deluxetable}
 
%---------------------------------
\subsection{Simulation parameters}
%---------------------------------

We have used two different $N$-body codes to carry out our simulations:
the Adaptive Refinement Tree code (ART; Kravtsov et al. 1997) and the
AP$^3$M code\footnote{The original public code was modified slightly to
  take into account $\ome\neq 1$ models.} (\cite{couchman91}).
Comparison of the results obtained with different numerical codes
allows us to insure the results are robust. AP$^3$M code is an
extension of the well-known P$^3$M algorithm (Hockney \& Eastwood
1981). The code performs hierarchical rectangular refinements in the
high-density regions to reduce expensive particle-particle
calculations. The gravitational force is obtained by matching the
gravitational forces calculated using FFT solver on the base and
refinement grids and the small-scale force calculated using direct
particle-particle summation (see Couchman 1991 for details). A total of
four refinement levels were allowed during the course of the \ap3m
simulations presented here. The ART code also reaches high force
resolution by refining all high-density regions with an automated
refinement algorithm.  The refinements are recursive: the refined
regions can also be refined, each subsequent refinement having half of
the previous level's cell size.  This creates an hierarchy of
refinement meshes of different resolution covering regions of interest.
The refinement is done cell-by-cell (individual cells can be refined or
de-refined) and meshes are not constrained to have a rectangular (or
any other) shape. This allows us to refine the required regions in an
efficient manner.  The criterion for refinement is {\em local
  overdensity} of particles: in the simulations presented in this paper
the code refined an individual cell only if the density of particles
(smoothed with the cloud-in-cell scheme; Hockney \& Eastwood 1981) was
higher than $n_{th}=5$ particles. Therefore, {\em all} regions with
overdensity higher than $\delta = n_{th}{\ }2^{3L}/\bar{n}$, where
$\bar{n}$ is the average number density of particles in the cube, were
refined to the refinement level $L$. For the two ART simulations
presented here, \lcdmxx and \lcdm30art, $\bar{n}$ is $1/8$ and $1$,
respectively.  The Poisson equation on the hierarchy of meshes is
solved first on the base grid and then on the subsequent refinement
levels. On each refinement level the code obtaines potential by solving
a Dirichlet boundary problem with boundary conditions provided by the
already existing solution at the previous level. There is no
particle-particle summation in the ART code.  The detailed description
of the code is given in Kravtsov et al.  (1997). Note, however, that
the present version of the code uses multiple time steps on different
refinement levels, as opposed to the constant time stepping in the
original version of the code. The multiple time stepping scheme is
described in some detail in Kravtsov et al. (1998; also see below).

The information about the numerical parameters of the simulations is
given in Table 2. The AP$^3$M code was used to produce four simulations
of different cosmological models\footnote{Runs SCDM, \lcdmtt, and
  \tCDM\ have not been completed due to the high computational
  expense. The simulations were stopped at $z=0.3$.}
 with the box size $L_{box} = 30\mpch$.
The size of the box side is a compromise between requirements of the
high spatial resolution ($\sim 2-4h^{-1} {\rm kpc}$) and good
statistics of halos. Nevertheless, for our most realistic model
($\Lambda$CDM) we also use the ART code to simulate a $60\mpch$ box
(\lcdmxx run).  We estimate effects of the finite box size (see \S~5.1)
on our results by comparing results of $140\mpch$ box simulation of
\cite{jenkins97} with results of our $60\mpch$ and $30\mpch$ boxes.
All \ap3m runs were done with $128^3$ particles.  The two ART runs used
$256^3$ particles.  The initial conditions were set using the
Zel'dovich approximation on uniform 128$^3$, $256^3$, and $512^3$
meshes for \ap3m runs, \lcdm30art~ run, and \lcdmxx run, respectively.
The seed used to generate the Gaussian random density field was the
same in all of our \ap3m runs, but different for each of the two ART
runs.  All of the simulations are started at the moment of time when
the rms density fluctuations at the Nyquist wavelength $\lambda_{Nyq}$
are still linear: $\sigma(\lambda_{Nyq},z_i) \sim 0.1-0.2$.  All \ap3m
runs evolved during a period in which the linear growth factor
increased by a factor of 50. This explains different values of $z_{init}$
for different models in Table 2.

As was explained in \S 1, the purpose of our study was to compute
the correlation function and the bias accounting for {\em all} DM halos,
including those inside groups and clusters. To assure that halos do
survive in clusters the force resolution should be $\sim 1-3h^{-1} {\rm
  kpc}$ (Moore et al. 1996; KGKK). Furthermore, if we aim to study
galaxy-size halos, the mass resolution should be $\lesssim 10^9h^{-1}
{\rm M_{\odot}}$ to resolve galaxy-size halos ($M\gtrsim 10^{11}h^{-1}
{\rm M_{\odot}}$) with at least $\approx 100$ particles. The compromise
between these considerations and the computational expense determined
the force and mass resolution of our simulations (see Table 2).  The
ART code integrates the equations of motion in {\em comoving}
coordinates.   However, the refinement strategy of the ART code is designed
to effectively preserve the initial {\em physical} resolution of the
simulation (see below). The peak resolution is reached by creating a
refinement hierarchy with six levels of refinement. In the \ap3m runs
the force resolution $\eta$ (spline softening length) was kept constant
in comoving coordinates while fluctuations are still in the linear
regime and is then set to be constant in {\em physical} units. The
switch occurs at the moment when first galaxy-size halos start to collapse ($z\sim
5-10$) for our simulations.  We chose to maintain fixed comoving
resolution until it reaches $\sim 3 \kpch$ (physical) (at $\sim
5-10$). At later moments the resolution is fixed to this value in
physical coordinates (the exception is the OCDM model in which the
resolution was set to $4.7\kpch$ by mistake). The dynamical range
$L_{box}/\eta$ of the simulations implied by the force resolution is
$\approx 16,000$ ($32,000$ formal) for the ART runs and $6000-10000$
for the \ap3m runs. The dynamic range of the \ap3m runs is just enough
to keep the initial physical resolution ($\approx 2-5\kpch$). The ART
code integrates the evolution in comoving coordinates. Therefore, in
order to prevent degradation of force resolution in {\em physical}
coordinates, the dynamic range between the start and the end ($z=0$) of
the simulation should increase by $(1+z_i)$: i.e., for our simulations
reach $512\times(1+z_i)=15,872$. This is accomplished with the prompt
successive refinements during the simulations.

The time stepping of the \ap3m and ART codes is rather different.
First of all, the codes integrate the equation of motion using
different time variables: the time in the \ap3m code and the
expansion factor in the ART code.  
In the \ap3m runs the time step is {\em constant} and is {\em
  the same} for all particles. In the ART runs, as was noted above, the
particles residing on different refinement levels move with different
time steps. The particles on the same level, however, move with the
same step. The refinement of the time integration mimicks spatial
refinement and step for each subsequent refinement level is two times
smaller than the step on the previous level. The global time step hierarchy
is thus set by the step $\Delta a_0$ at the zeroth level (uniform base
grid). The step on level $L$ is then $\Delta a_L=\Delta a_0/2^L$.  The
choice of an appropriate time step for a simulation is dictated by the
adopted force resolution. The number of time steps in our simulations
is such that the {\it rms} displacement of particles during
a single time-step is always less than $\eta /4$ (less than 1/4 of a cell
in the ART code)\footnote{Note, however, that the distance traveled by
  the fastest moving particle in one time-step in \ap3m runs can be
  larger than $\eta$, especially at late times and in the \lcdmtt run.
  In the ART code particles do not move further than $\sim 0.5$ cells
  in a single time step, where the cell size and time step for particles
  located on the refinement level $L$ are $\Delta x_0/2^L$ and $\Delta
  a_0/2^L$, respectively.}. The size of the time-step, $\Delta t$, for
the \ap3m runs was chosen to be sufficiently small to satisfy the stability
criteria of the numerical integration (\eg\ \cite{edfw85}) throughout
the entire run. In the case of ART runs, the value of $\Delta
a_0=0.0015$ was determined in a convergence study using a set of
smaller $64^3$ particle simulations described in Kravtsov et
al. (1998). In both \ap3m and ART runs the energy was conserved
with an accuracy $\lesssim 1\%$.

\begin{deluxetable}{llcccccc}
%\tablewidth{58pc}
\tablecolumns{4}
\tablecaption{Parameters of Halo Finding Algorithm}
\tablehead{\colhead{Run} & \colhead{\delth} & \colhead{\mhalo} & \colhead{$r_s$} 
 \\ 
 & & (\msunh) & (kpc/h) }
\startdata
SCDM\phm{100} & 180 & $7\times 10^{10}$ & 13  \nl 
OCDM\phm{100} & 180-400 & $2\times 10^{10}$ & 13  \nl
\lcdmtt\phm{100} & 180-340 & $2\times 10^{10}$ & 13  \nl
\lcdmxx\phm{100} & 180-340 & $ 10^{10}$ & 20  \nl
\lcdm30art\phm{100} & 180-340 & $ 10^{9}$ & 10  \nl
\tCDM\phm{100} & 180 & $7\times 10^{10}$ & 13  \nl 
\enddata
\end{deluxetable}

%-----------------------------------
\subsection{Identification of halos}
%-----------------------------------

Identification of DM halos in the very high-density environments (e.g.,
inside groups and clusters) is a challenging problem.  Traditional halo
finding algorithms, such as friends-of-friends (e.g., Davis et al.
1985) or ``overdensity-200'' (e.g., Lacey \& Cole 1994), cannot be
used.  These algorithms are not designed to search for substructure;
they identify an isolated halo above virial overdensity as a single
object and cannot account for the internal substructure. Our goal,
however, is to identify both isolated halos and halos orbiting within
larger systems (``sub-halos''). The problems associated with halo
identification within high-density regions are discussed in KGKK. In
this study we use a halo finding algorithm called Bound Density Maxima
(BDM; see KGKK). A detailed description of the working version of the
BDM algorithm used here can be found in Klypin \& Holtzman (1997).
Other recently developed algorithms capable of identifying satellite
halos are described in Ghigna et al. (1998) and KGKK.  The main idea of
the BDM algorithm is to find positions of local maxima in the density
field smoothed at a certain scale and to apply physically motivated
criteria to test whether the identified site corresponds to a
gravitationally bound halo. In the following we describe specific
parameters of the BDM used to construct halo catalogs used in our study
(the main parameters are listed in Table 3).

The radius of a halo assigned to it by the algorithm is either its
virial radius\footnote{The virial overdenisty $\delta_{TH}$ is set in
  accord with the prediction of the top-hat collapse model. Note that
  $\delta_{TH}$ depends on the cosmological model (e.g., Kitayama \&
  Suto 1996).} or $150\kpch$, whichever is smaller. The latter is
approximately the maximum virial radius we would expect for a
galaxy-size halo. The mass and radius are very poorly defined for the
satellite halos due to the tidal stripping which alters a halo's mass
and physical extent. Therefore, in this study we will use maximum
circular velocity $V_{max}$ as a proxy for halo mass. This allows us to
avoid complications related to the mass and radius determination for
satellite halos. For isolated halos $V_{max}$ and the halo's virial
mass are directly related. For example, for a halo with a density
distribution described by the Navarro, Frenk \& White (hereafter NFW;
1996) profile $\rho(r)\propto x^{-1}(1+x)^{-2}$ ($x\equiv r/R_s$; $R_s$
is scale-radius):
\begin{equation}
V_{max}^2=\frac{GM_{vir}}{R_{vir}}\frac{c}{f(c)}\frac{f(2)}{2};
\end{equation}
where $M_{vir}$ and $R_{vir}$ are virial mass and radius, $f(x)\equiv
\ln(1+x)-x/(1+x)$, $c\equiv R_{vir}/R_s$.  While for the
sub-halos $V_{max}$ may not be related to mass in any obvious way, it
is still the most physically and observationally motivated halo quantity.
The limiting radius of $150\kpch$ is sufficient to determine the $V_{max}$
for galaxy-size halos. The cluster-size halos are not explicitly
excluded from the halo catalogs. We assume therefore that the center of
each cluster can be associated with a central cluster galaxy. The
latter (due to the lack of hydrodynamics and other relevant processes)
cannot be identified in our simulations in any other way.

The density maxima are identified using a top-hat filter with radius
$r_s$ (``search radius''). The search is performed starting from a
large number of randomly placed positions (``seeds'') and proceeds by
moving the center of mass within a sphere of radius $r_s$ iteratively
until convergence. In order to make sure that we use a sufficiently
large number of seeds, we used the position of every tenth particle as a
seed. Therefore, the number of seeds by far exceeds the number of
expected halos.  The search radius $r_s$ also defines the minimum
allowed distance between two halos. If the distance between centers of
any of the two halos is $<2r_s$, only one halo (the more massive of the
two) is left in the catalog. A typical value for the search radius is
$(10-20)\kpch$. We set a lower limit for the mass inside the search
radius $M(<r_s)$: halos with $M(<r_s)<\mhalo$ are not included in the
catalog. This is done to exclude pure poisson fluctuations from the
list of halo candidates.  Some halos may have significant substructure
in their cores due, for example, to an incomplete merger. Such cases
appear in the catalogs as multiple (2-3) halos with very similar
properties (mass, velocity, radius) at small separations.  Our strategy
is to count these as a single halo. Specific criteria used to identify
such cases are: (1) distance between halo centers is $\lesssim
150\kpch$, (2) their relative velocity in units of the {rms} velocity
of particles in the halos $\Delta v/v$ is less than 0.15, and (3) the
difference in mass is less than factor 1.5. Only the most massive halo
is kept in the catalog.

It is obvious that for a statistical study it is important to be
confident that our BDM algorithm does not miss a large fraction of
halos. While a substantial effort is made to reject fake halos, it is
important to make sure that all real halos are included in the catalog.
The algorithm is very efficient at finding isolated halos and the major
difficulties are in identification of halos in crowded regions. Halo
identification in such regions is complicated by the large number of
halos and by the high-density background of fast moving particles.
Therefore, performance of the algorithm in such regions is a good
indicator of its overall performance. In fact, the parameters of the
halo finder used to construct our halo catalogs were tuned by visual
inspections of the most difficult and complicated regions.  An example
of such regions is shown in Figure 1. The figure shows the distribution of
the DM particles in a Virgo-type cluster\footnote{The distribution of
  matter and halos in this cluster is discussed in the next section.}
in the \LCDM\ model (\lcdmxx run). To enhance the substructure,
particles are color-coded on a grey-scale with a rendering algorithm
based on the local particle density (see caption). A large number of
distinct and compact DM halos is clearly present within the virial
radius of the cluster.  Figure 2 shows positions of DM halos identified
by the BDM code in the same volume.  There are 121 halos in the plot,
each halo having more than 20 bound particles. More stringent limits of
more than 30 particles and a maximum circular velocity, $V_{max}$, larger
than $80\kms$ produce 98 halos. All distinct halos visible in Figure 1,
are identified.

\begin{figure*}[ht]
\pspicture(0,8.0)(15.0,22.6)
%\psgrid(0,8)(18.8,20)
\rput[tl]{0}(-2.0,22.8){\epsfysize=11.5cm
\epsffile{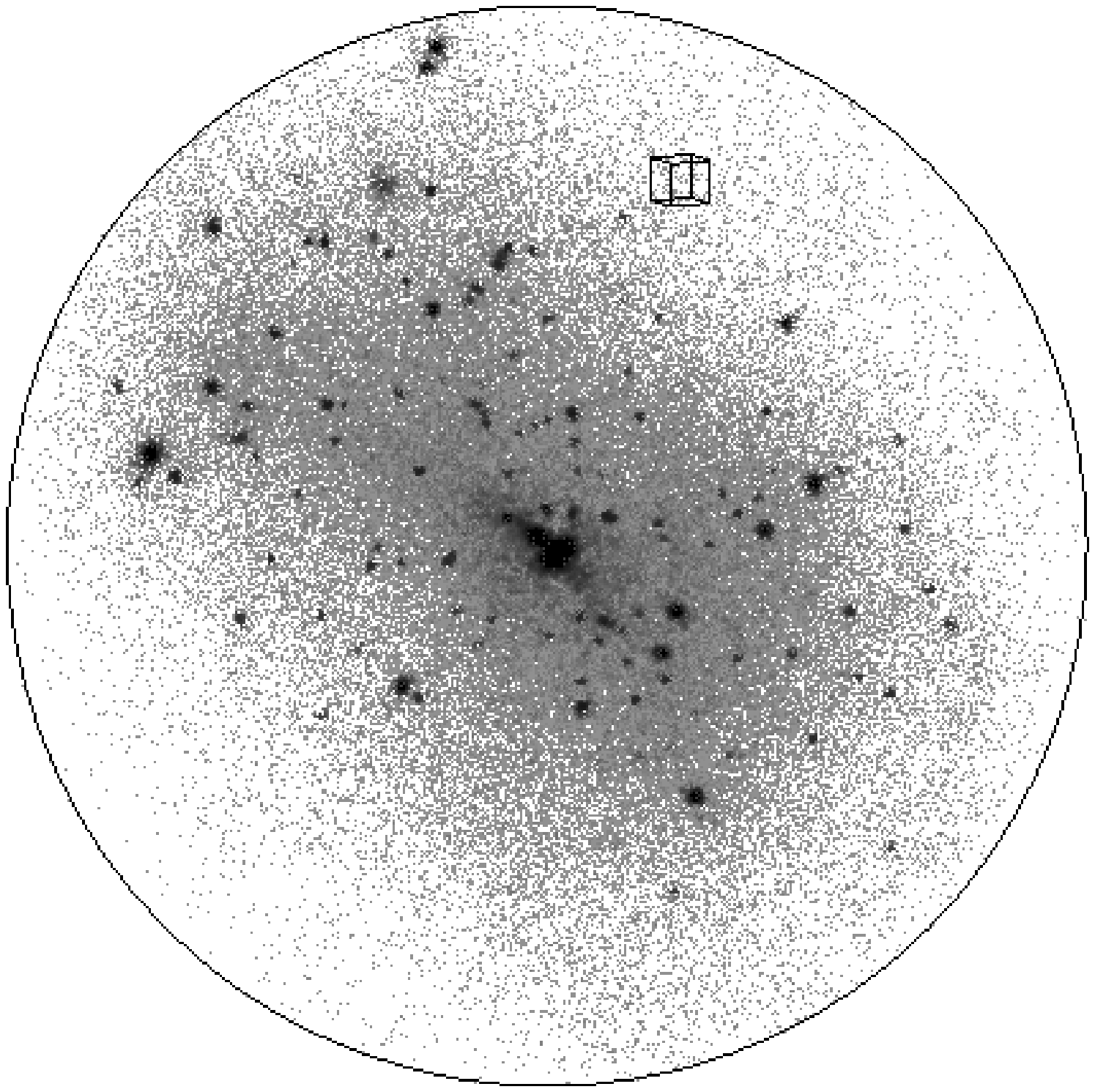}}
\rput[tl]{0}(8.5,22.8){\epsfysize=11.5cm
\epsffile{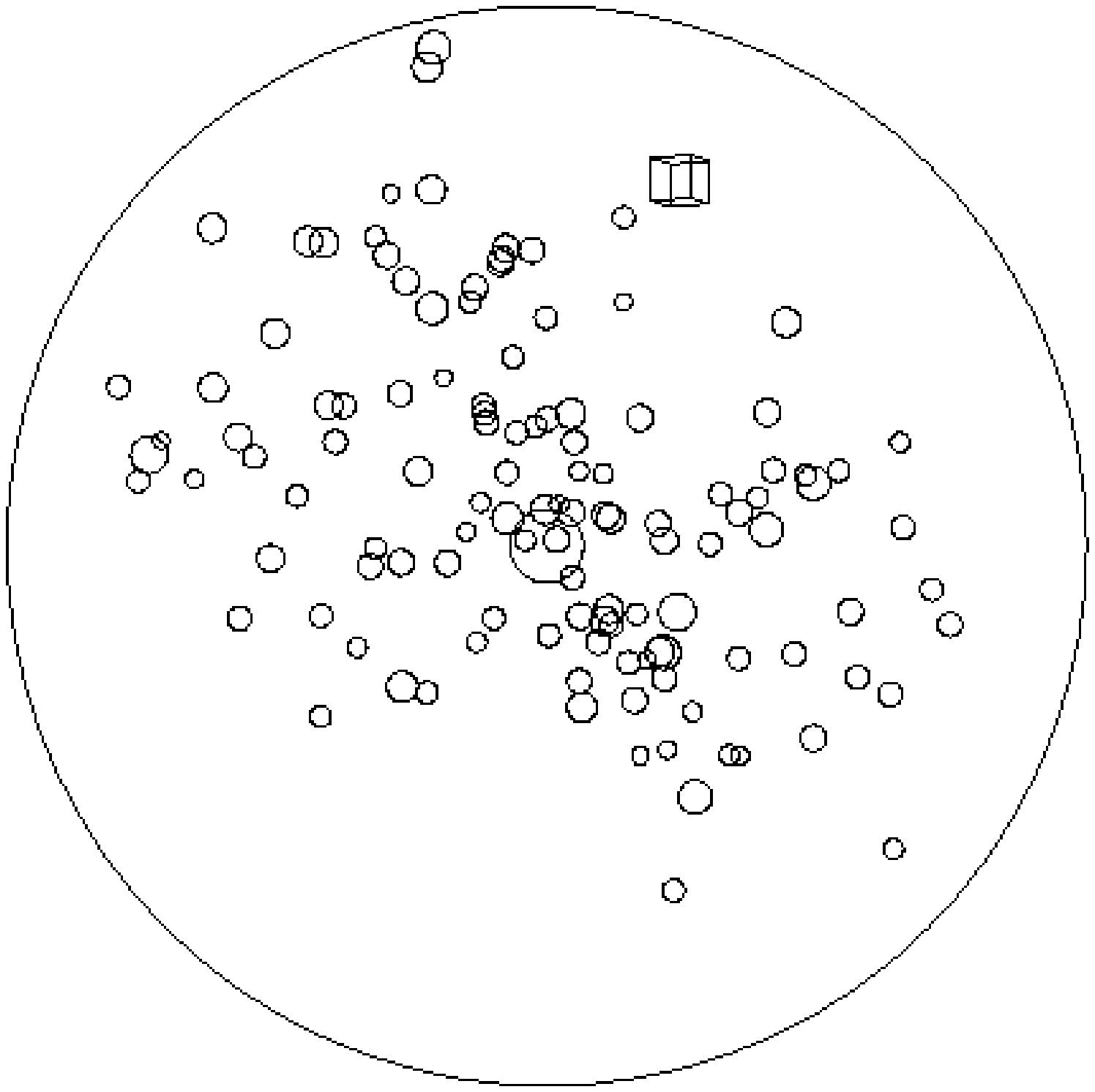}}
\rput[tl]{0}(0.0,11.3){
\begin{minipage}{8.9cm}
  \small\parindent=3.5mm {\sc Fig.}~1.---
   Distribution of dark matter in a Virgo-like cluster in the
  \LCDM$_{60}$ simulations. The cluster virial mass is $2.45\times
  10^{14}\msunh$ and the corresponding 3D velocity dispersion is
  $\approx 1022\kms$. The particles inside a sphere of the radius of
  1.5\mpch~(solid circle) are shown. The size of the small box, shown
  to provide the comparison scale, is 100\kpch. To enhance the
  contrast, we have color-coded DM particles on a grey scale according
  to their local density: intensity of each particle is scaled as the
  logarithm of the density difference $\rho_{15}-\rho_{75}$, where the
  densities were obtained using top-hat filter with radii 15\kpch~and
  75\kpch.
\end{minipage}
}
\rput[tl]{0}(9.6,11.3){
\begin{minipage}{8.9cm}
  \small\parindent=3.5mm {\sc Fig.}~2.--- The region of the space shown
  in Figure 1 with circles representing the DM halos found by the BDM
  halo finder.  Area of each circle representing a halo is proportional
  to the halo's maximum circular velocity. There are 121 halos in the
  plot, each of which has mass $>2\times 10^{10}h^{-1} {\rm M_{\odot}}$
  (more than 20 bound particles).
\end{minipage}
}
\endpspicture
\end{figure*}

%-----------------------------------------
\subsection{Survival of halos in clusters}
%-----------------------------------------

In \S~1 we have stressed that the main new feature of the analysis
presented in this paper is the identification and use of halos located
inside the virial radius of other halos (satellites or sub-halos).
Particularly, one of the main goals is to include halos inside
cluster-size halos. This would allow an estimation of the halo-halo correlation
function down to unprecedentedly small scales ($\sim 150\kpch$).
However, we can consider this estimate robust only if we are confident
that no halos (or a very small fraction of them) are artificially
destroyed in clusters due to the insufficient resolution. KGKK have
discussed analytical estimates and numerical experiments that could be
used to address this issue.  Following the approach of KGKK, we have
run a series of small $N$-body simulations\footnote{similar to those of
  KGKK but with all parameters appropriate for the $\Lambda$CDM model.}
using the direct-summation Aarseth's code (Binney \& Tremain 1987). The
basic setup of the simulations is as follows. A DM halo of virial mass
$10^{12}\msunh$ (containing a few thousand particles within the virial
radius) is constructed. The initial equilibrium density profile of the
halo is described by the NFW formula. The halo is then placed on an
orbit in a constant potential corresponding to a galaxy cluster of mass
$2\times 10^{14}$\msunh with the NFW density distribution.  The
particular numbers quoted here are intended to mimic the orbital
evolution in the cluster presented in Figs. 1-3. The orbital
evolution of the halo was studied for different orbits and different
(mass and force) resolutions. Both the mass and the force resolution
were varied by more than a factor of ten.  The orbital evolution of
mass bound to the halo converges at the force resolution\footnote{Here
  we quote all resolutions for the spline kernel $\eta$ of the \ap3m,
  the corresponding resolution is $\approx \eta/2$ for the ART.}
$\approx 3-4\kpch$: i.e., the evolution of bound mass in runs with
higher resolution is {\em identical}. In fact, during the first
$\approx 5$ Gyrs the resolution of $\approx 10\kpch$ is adequate. The
mass loss in this case is somewhat higher which leads to total
destruction after 5 Gyrs, whereas resolution of $3\kpch$ allows halo to
survive during the Hubble time.  These experiments have also shown that
mass resolution of $\approx 10^9\msunh$ is sufficient, provided that
force resolution is high. Two runs with force resolution of $3\kpch$
and with mass resolutions of $10^8\msunh$ and $10^9\msunh$ resulted in
identical mass evolution. It is worth noting that in these experiments
the halo was followed until it was totally destroyed by the tidal field
($\gtrsim 5$ Gyrs in most cases). In real simulations, however,
clusters form only at $z\lesssim 1$, and most of the accreted halos
spend $\lesssim 5$ Gyrs in clusters. During this time halos lose
80-90\% (depending on the orbit) of their initial mass. Thus, if the mass
resolution is $\approx 10^9\msunh$, the halos with initial mass
$\gtrsim 10^{11}\msunh$ can be identified even after spending a
substantial time in a cluster.

Details of the evolution of halos in clusters depend sensitively on the
parameters of the halo orbit. Specifically, the mass loss rate depends
on the pericenter and on the eccentricity of the orbit. A halo on a
very eccentric orbit survives for a considerably longer time than a
halo on a circular orbit, even if the radius of the latter is larger than
the pericenter. For example, a halo on a circular orbit with a radius
of $250\kpch$ was {\it totally destroyed} in less than 5 Gyrs, 
while a halo on a very eccentric orbit with the pericenter of
only $125\kpch$ {\it survived} for more than 10 Gyrs. The
explanation for this is simple: a halo on a circular orbit spends all of
its time in a high-density cluster core suffering a steady mass loss,
while a halo on the eccentric orbit spends only a small
fraction of its orbital period in the core.

With the resolution quoted above, only a relatively small fraction of
halos can be tidally destroyed in clusters. Only the halos with the
{\em apocenter} $\lesssim 300\kpch$ are subject to the destruction. It
can be expected that halos which are accreted when the cluster was young
and its radius was small ($\sim 300-500\kpch$) have small apocenters. 
The fraction of such halos in a $z=0$ cluster is small. Our estimates
show that for halos with large apocenters the dynamical friction cannot
bring the apocenter considerably closer to the cluster center.  The
tidal stripping reduces the halo mass very efficiently, thus increasing
the friction time.

The tidal stripping that halos suffer in clusters has an important
effect on their density profiles. It appears that the halo profile is
affected at all radii and not only at radii close to the tidal radius.
Indeed, the trajectories of particles in a halo are mostly eccentric
and after a crossing time the absence of the stripped particles will be
``felt'' by the whole halo.  The change in the density profile means,
of course, a change in the maximum circular velocity of the halo.  The
effect is not dramatic because the fraction of particles on highly
eccentric orbits is rather small and the central regions of halos are
thus affected the least. Typically, the circular velocity of the halos is
reduced during their evolution in the cluster by 20\%--30\%. This
correction was taken into account in Figure 3 by reducing the velocity cut 
from 120 km/sec at z=1.0 to 100 km/sec at z=0.0.

The tidal destruction is important only for cluster-size
halos (where tidal fields are the strongest). There are three massive
clusters in our \lcdmxx simulation with mass $\gtrsim 2\times
10^{14}\msunh$. There is one ``Coma cluster'' with velocity
dispersion $\sigma_{3D}=1654{\ }\kms$ and mass $M=6.4\times 10^{14}\msunh$
inside $1.43\mpch$ radius.  The cluster contains 201 halos of mass
$>3\times 10^{10}\msunh$ inside a 2\mpch\ box and 9 halos inside
0.3\mpch.  Unfortunately, the cluster has suffered a recent major
merger and it was difficult to study the radial distribution of halos.
One of the other two clusters also shows an indication of an ongoing
merger. It has two sub-clusters in the central region separated by
$\approx 0.5\mpch$. Therefore, we will focus on the third cluster which
has a relatively regular appearance (see Fig.1).  

The cluster contains 121 bound halos with $\gtrsim 20$ particles inside
1.5\mpch. There are 231,200 dark matter particles inside the virial radius
of 1.28\mpch. The virial mass of the cluster has increased from $M_{\rm
  vir}=7.9\times 10^{13}\msunh$ at $z=1$ to $M_{\rm vir}=2.4\times
10^{14}\msunh$ at $z=0$.  Figure 3 shows the density profile of matter
and the radial number density profile of DM halos inside the cluster.
The DM density profile at $z=0$ is well approximated by the NFW
profile.  For comparison, we also present the density profile of this
cluster at redshift one. Note that at $z=1$ the radius is given in
proper units and the mean matter density ($\rho_0$) is estimated at
$z=0$.  Similarly, the bottom panel of Figure 3 shows the number
density profile of halos in the cluster at $z=0$ and at $z=1$.  Halos
with more than 30 {\em bound} particles and with limits on the maximum
circular velocity $V_{max}\geq 100\kms$ and $V_{max}\geq 120\kms$ (the
change is for the reasons explained above) for $z=0$ and $z=1$,
respectively, were used. The mean number density of halos, $n_0$, was
estimated using all halos in the simulation within the above velocity
limits ($N_h=7628$ at $z=0$).  The profile at $z=1$ is also rescaled
into proper units.  Figure 3 clearly indicates that the number of halos
in the central $300\kpch$ (proper radius) has declined substantially
from $z=1$ to the present epoch. In the central $300\kpch$ there are
three times as many halos (24) at $z=1.0$ than at $z=0$.  If we interpret
the difference as due to tidal destruction, then 16 halos were
destroyed in the central part. The situation is different at larger
radii: there are 57 and 50 halos at $0.3<r<1.3\mpch$ for $z=1$ and
$z=0$, respectively. Note, that we compare the number of halos in the {\em
  same proper volume} at these two moments: the volume corresponding to
the virial radius of the cluster at $z=0$ (the virial radius at $z=1$
is smaller).  The number density profile is thus virtually unaffected
at $r\gtrsim 300$\kpch.

%=================

\section{Results}

%=================

In this section we will present results on the evolution of the 2-point
correlation function and bias (as defined by eq. [1]) in the
simulations described in the previous section. The presentation of the
results is split in two subsections. In the first of them we present
the results of our largest simulation: $\Lambda$CDM$_{60}$. In the
second section we present the results of the rest of our simulations of
different cosmological models. Thus, we will first focus on the results
of the most realistic of the studied models, \LCDM\ (see \S~3), and
then will discuss the differences between cosmological models. However,
before we proceed with the presentation of the results, we will first
compare our estimate of the dark matter correlation function with
estimates which have been performed by different authors 
{\pspicture(0.5,-1.5)(13.0,17.0)
\rput[tl]{0}(-1.0,16.7){\epsfxsize=12cm
\epsffile{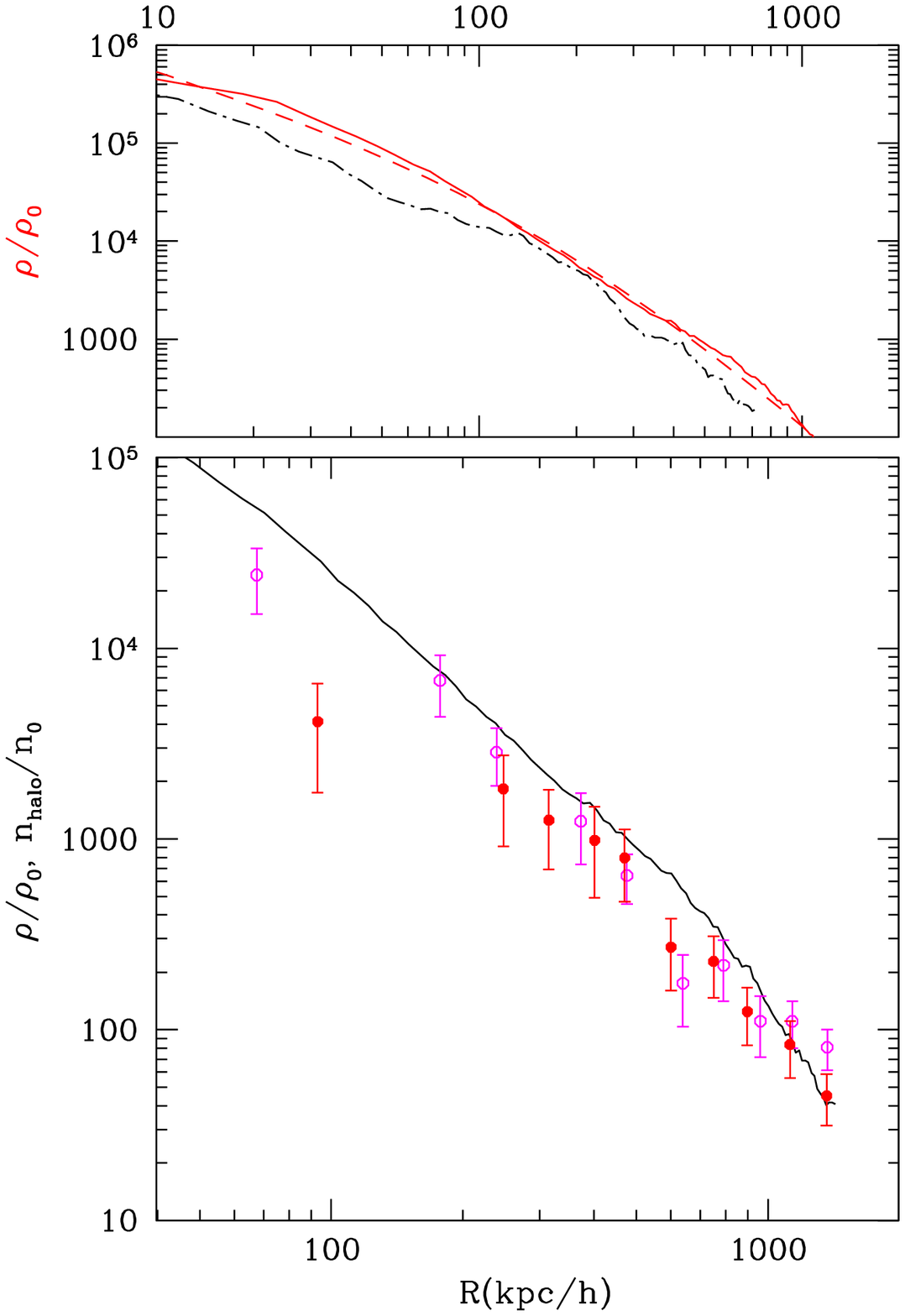}}
\rput[tl]{0}(0.4,4.5){
\begin{minipage}{8.7cm}
  \small\parindent=3.5mm {\sc Fig.}~3.--- The density profile of the
  cluster shown in Figures 1 and 2. {\it Top panel:} Dark matter
  density in units of the mean matter density at $z=0$ ({\em solid
    line}) and at $z=1$ ({\em dot-dashed line}); the {\em dashed line}
  shows the Navarro, Frenk \& White (1996) fit to the $z=0$ density
  profile.  The $z=1$ density profile is given in {\em proper} units:
  the radius is given in proper scale and the mean matter density is
  estimated at $z=0$. {\it Bottom panel:} Number density profiles of
  halos in the cluster at $z=0$ ({\em solid circles}) and at $z=1$
  ({\em open circles}) as compared to the $z=0$ density profile of dark
  matter ({\em solid curve}).  The error-bars show $1\sigma$ poisson
  errors. Halos with more than 30 bound particles and with maximum
  circular velocity larger than $100\kms$ at $z=0$ and larger than
  $120\kms$ at $z=1$ were used to estimate the average density $n_0$
  (see \S~4.3 for details). The profile at $z=1$ is rescaled into
  proper units similarly to that of the dark matter in the top panel.
\end{minipage}
}
\endpspicture}
and with different
numerical codes.

%-------------------------------------------------------------------
\subsection{The dark matter correlation function in the $\Lambda$CDM
  model: comparison with other studies}
%-------------------------------------------------------------------

In Figure 4 we compare the 2-point correlation function of
the {\em dark matter} in the \LCDM\ model (the ART run
$\Lambda$CDM$_{60}$) with similar estimates presented by Klypin et al.
 (1996) and by Jenkins et al. (1998). The two 
latter estimates were done with different codes (PM in Klypin et al.;
and AP$^3$M in Jenkins et al.) and different resolutions (cell size of
$62\kpch$ for the PM, and $30\kpch$ (Plummer) for the AP$^3$M). All
simulations followed $256^3$ particles, although physical mass
resolution was different due to the different box sizes
($50\mpch$, $60\mpch$, and $141.3\mpch$ for the PM, ART, and \ap3m
simulations). 
{\pspicture(0.5,-1.5)(13.0,13.0)
\rput[tl]{0}(-0.5,13.3){\epsfxsize=10.5cm
\epsffile{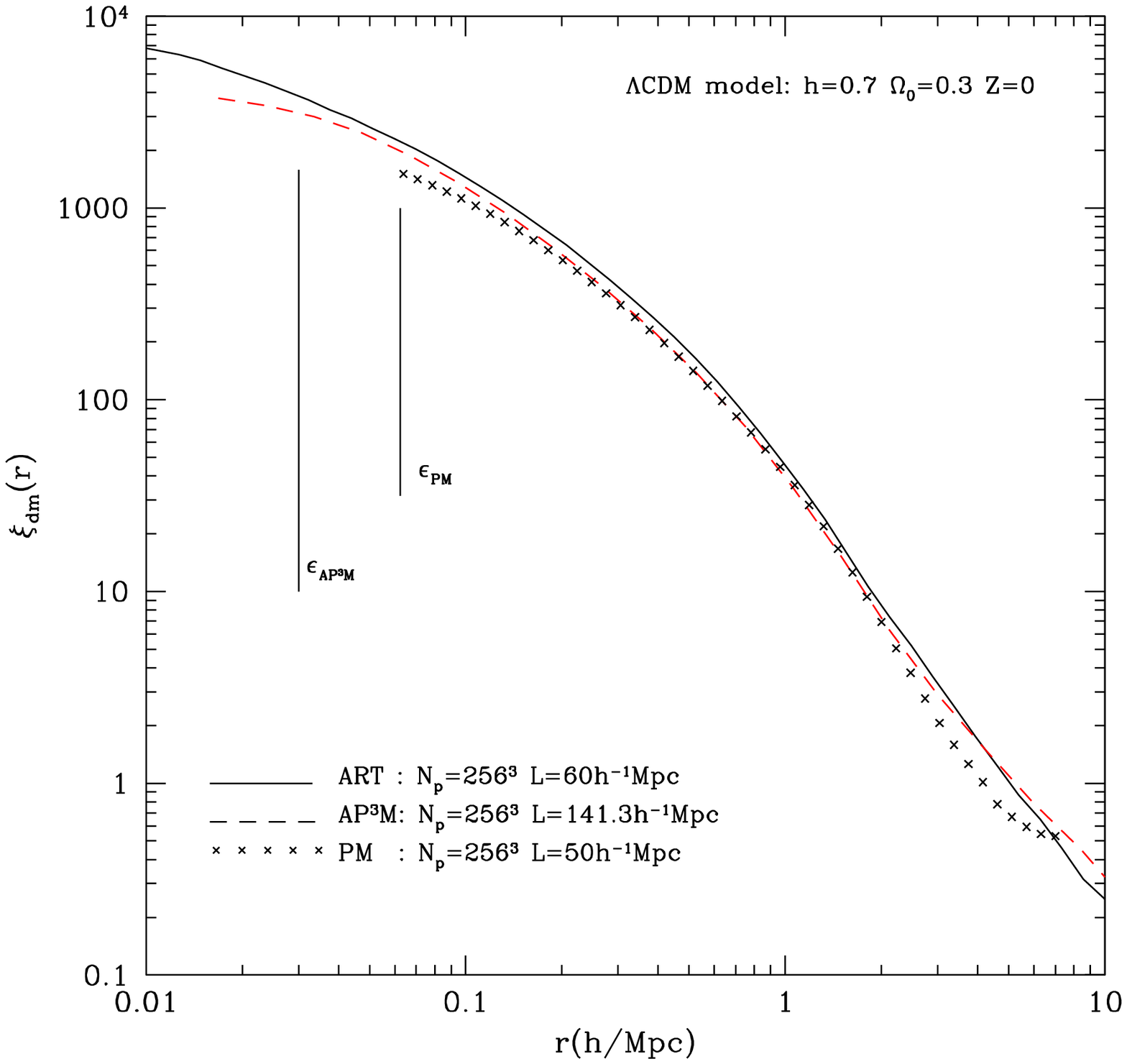}}
\rput[tl]{0}(0.4,3.5){
\begin{minipage}{8.7cm}
  \small\parindent=3.5mm {\sc Fig.}~4.--- The comparison of the
  correlation functions of the dark matter {\xidm} in the \LCDM~ model
  estimated by different authors with different numerical resolutions
  and codes.  The {\em solid curve} shows {\xidm} in our \LCDM$_{60}$
  run, simulated using the ART code.  The {\em dashed curve} shows
  {\xidm} estimated by Jenkins et al. (1998) using the \ap3m code.  The
  {\em crosses} show {\xidm} estimated by Klypin et al. (1996) using
  the PM code.  The {\ap3m} and PM correlation functions were rescaled
  as described in \S~5.1 to account for the difference in normalization
  between different simulations. All simulations followed evolution of
  $256^3$ particles.  The vertical lines indicate formal force
  resolution for each code (the line for the ART code at $1.8\kpch$ is
  off the plot).
\end{minipage}
}
\endpspicture}
 Although the cosmological model was exactly the same in
all three estimates\footnote{Note, however, that there are small
  differences in the approximation used for the power spectrum between
  the simulations.}, the normalization of the power spectrum was
slightly different. The rms mass fluctuations on scale of $8\mpch$
$\sigma_8$ was 0.9, 1.0, and 1.1 for the AP$^3$M, ART, and PM
simulations, respectively.  Therefore, we have multiplied (divided) the
DM correlation function of AP$^3$M (PM) simulation by $1.1^2$ in order
to account for the differences in normalization.  Figure 4 indicates
that there is a very good agreement between all estimates at scales
$\approx (0.2-2)\mpch$. The ART and \ap3m estimates agree to better
than 10\% at scales $0.03-7\mpch$!  On larger scales the ART correlation
function has a smaller amplitude than that of the \ap3m due to a factor
of 2.35 smaller box size of the former. This result is in agreement with
previous studies (e.g., Col\'in et al. 1997 and references therein)
which concluded that the correlation function is underestimated at scales
$\gtrsim 0.1$ of the simulation box size. Nevertheless, the agreement
is striking at smaller scales, given all the differences ({\em
  including cosmic variance}) between the estimates. We conclude,
therefore, that the correlation functions presented in the next
subsection are robust at scales $\lesssim 7\mpch$. This scale is, of
course, lower ($\approx 3-4\mpch$) for the $30\mpch$ simulations presented
in \S~5.3.

\begin{deluxetable}{ccccccccc}
%\scriptsize
\tablecolumns{13}
\tablewidth{0pc}
 \tablecaption{Power-law fits to $\xi_{hh}$ in the $\Lambda$CDM$_{60}$
   simulation}
\tablehead{
\\
\colhead{}    & 
\multicolumn{2}{c}{$V_{max}>120 {\rm km/s}$} &\colhead{}    &
 \multicolumn{2}{c}{$V_{max}>150 {\rm km/s}$} & \colhead{}    &
\multicolumn{2}{c}{$V_{max}>200 {\rm km/s}$}     \\ \\
\cline{2-3} \cline{5-6} \cline{8-9}\\
\colhead{z} & \colhead{$r_0$} & \colhead{$\gamma$} &\colhead{}    &
\colhead{$r_0$} & \colhead{$\gamma$} &\colhead{}    &
 \colhead{$r_0$} & \colhead{$\gamma$} 
    }
\startdata
0.0 & 4.864$\pm$0.011 & 1.704$\pm$0.004 & 
    & 4.789$\pm$0.024 & 1.687$\pm$0.009 & & 5.082$\pm$0.062 & 1.650$\pm$0.022\nl  
0.5 & 4.145$\pm$0.009 & 1.684$\pm$0.004 & 
    & 4.040$\pm$0.019 & 1.726$\pm$0.009 & & 4.534$\pm$0.049 & 1.777$\pm$0.020\nl 
1.0 & 3.760$\pm$0.007 & 1.694$\pm$0.003 & 
    & 3.944$\pm$0.015 & 1.762$\pm$0.007 & & 4.387$\pm$0.038 & 1.869$\pm$0.016\nl  
1.5 & 3.419$\pm$0.006 & 1.642$\pm$0.003 &  
    & 3.823$\pm$0.013 & 1.715$\pm$0.006 & & 4.288$\pm$0.035 & 1.831$\pm$0.016\nl
2.0 & 3.103$\pm$0.006 & 1.588$\pm$0.003 & 
    & 3.831$\pm$0.011 & 1.629$\pm$0.006 & & 4.375$\pm$0.036 & 1.702$\pm$0.016\nl 
3.0 & 2.974$\pm$0.006 & 1.527$\pm$0.004 & 
    & 3.453$\pm$0.011 & 1.536$\pm$0.006 & & 4.327$\pm$0.037 & 1.683$\pm$0.016\nl
4.0 & 3.206$\pm$0.009 & 1.516$\pm$0.005 & 
    & 3.356$\pm$0.011 & 1.498$\pm$0.006 & & 4.733$\pm$0.044 & 1.598$\pm$0.017\nl
5.0 & 3.602$\pm$0.013 & 1.534$\pm$0.007 & 
    & 3.661$\pm$0.016 & 1.516$\pm$0.008 & & 4.829$\pm$0.053 & 1.578$\pm$0.020\nl
\tablenotetext{a}
 { Column description: $z$ is redshift; $r_0$ and $\gamma$ are the best
   fit parameters of the power-law $\xi_{hh}(r)=(r/r_0)^{-\gamma}$ fit
   to the {\em comoving} halo-halo correlation function; $r_0$ is {\em
     comoving} in units $h^{-1} {\rm Mpc}$.
 }
\enddata
\end{deluxetable}
{\pspicture(0.5,-1.5)(13.0,15.0)
\rput[tl]{0}(-2.3,14.7){\epsfxsize=13cm
\epsffile{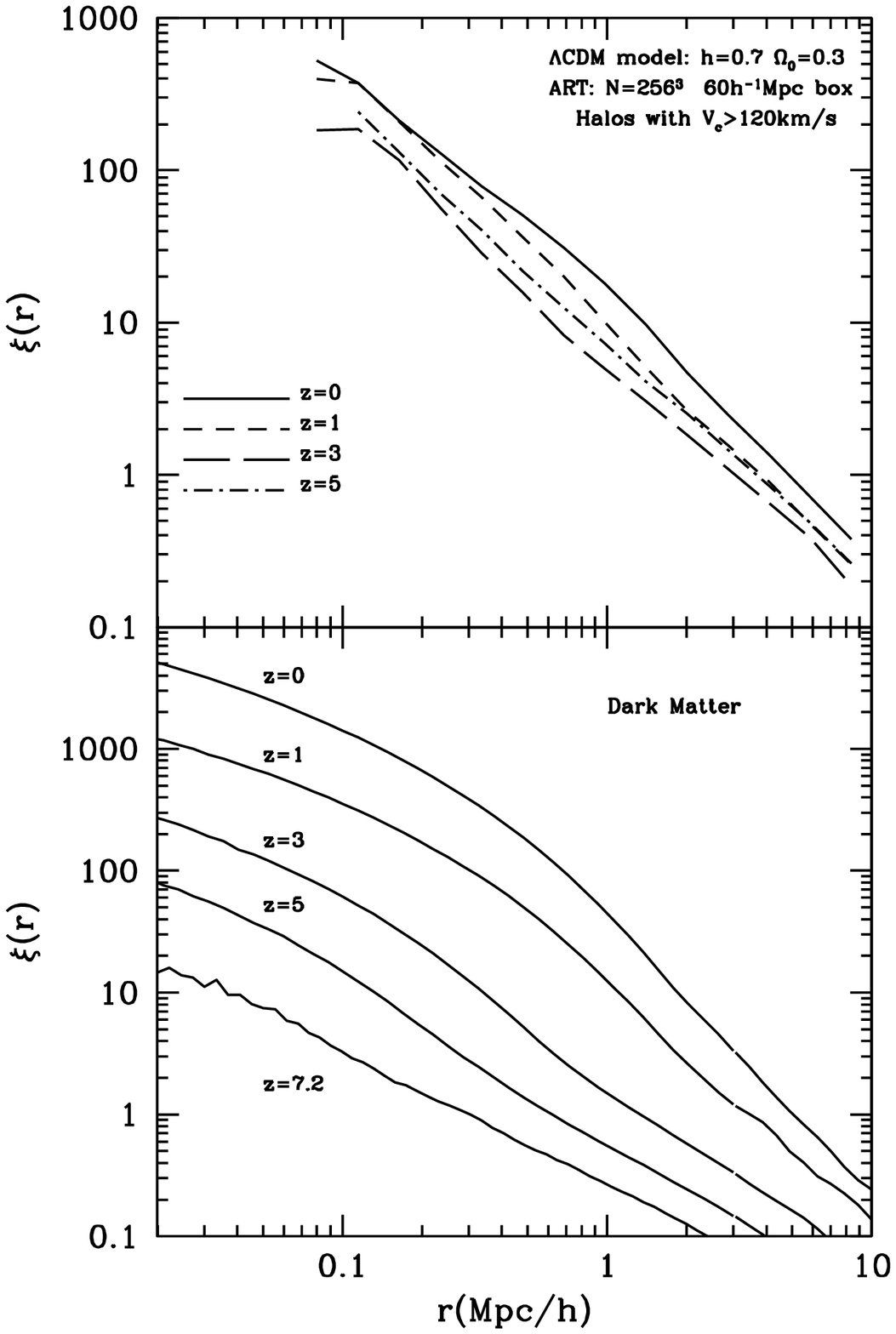}}
\rput[tl]{0}(0.0,1.5){
\begin{minipage}{8.7cm}
  \small\parindent=3.5mm {\sc Fig.}~5.--- The evolution of the 2-point
  correlation function of the dark matter (bottom panel) and halos (top
  panel) in the \LCDM$_{60}$ simulation. Only halos with maximum
  circular velocity $> 120\kms$ were used to estimate the halo
  correlation functions.  Poisson errors for the halo correlation
  functions are negligible at scales $\gtrsim 0.2\mpch$ and are not
  shown for clarity; at scales $<0.2\mpch$ the errorbars are $\lesssim
  20\%$ (see \S~5.2 for details). The best fit parameters of the
  power-law fit to the halo correlation functions are given in Table 4.
\end{minipage}
}
\endpspicture}
%------------------------------------------------------------
\subsection{Evolution of the correlation function and bias
  in the \LCDM\ model}
%------------------------------------------------------------

In Figure 5 we plot the evolution of the correlation function of both the 
dark matter $\xi_{dm}$ and the DM halos $\xi_{hh}$ in our \lcdmxx 
simulation. All results in this and the following sections are presented in {\em
  comoving} coordinates.  The halo-halo correlation function was
constructed using halos with $V_{max}\geq 120\kms$. Note that although
the earliest moment at which we show the correlation function is $z=5$,
the first halos in this simulation collapsed at redshift $z\approx 10$
and hundreds of halos are identified at $z\approx 7-8$. The number of
halos in the $V_{max}\geq 120\kms$ catalog is approximately $4300$,
$10,000$, and $7500$ for redshifts of 5, 2, and 0, respectively. The
good statistics result in a very accurate estimate of the correlation
function. The number of pairs per scale bin used to estimate $\xi_{hh}$
is $>50$ in all cases. Typically, $60-150$ for $r<0.2\mpch$, $200-1000$
for $0.2<r<1\mpch$, and $>1000$ for the larger scales. The pure poisson
errors associated with each of the points are thus neglegibly small,
except for the first 2-3 bins (where poisson errors are still small: $\lesssim
20\%$). There are subtler errors associated with radial binning, but
these are also less than a few per cent.

Figure 5 shows that the shapes of the matter and halo-halo correlation
functions are quite different. The matter correlation function changes
its shape from almost a power-law to a complicated shape. The slope of
$\xi_{dm}$ at scales $\lesssim 0.5\mpch$ stays approximately constant
throughout the evolution, while at the larger scales $\xi_{dm}$
significantly {\em steepens}. The amplitude of the DM correlation
function increases from $z=5$ to $z=0$ by factors of $\approx 60$ and
$\approx 10$ at small and large scales, respectively.  The halo-halo
correlation function behaves very differently. Its shape can be
well-described by a power-law at all epochs. The amplitude of
$\xi_{hh}$ evolves non-monotonically: it {\em decreases} somewhat from
$z=5$ to $z=3$, and then gradually increases. The evolution, however,
is much more modest than the dramatic evolution of the $\xi_{dm}$: the
maximum difference in amplitude among any two epochs is only a factor
of two. The details of the $\xi_{hh}$ evolution are illustrated in
Figure 6. This figure shows the evolution of the $\xi_{hh}$ amplitude
at a variety of different {\em comoving} scales (indicated on the
right) for the three {\LCDM} runs from our set of simulations (see
Table 2).  Different initial conditions of these runs allow us to
evaluate the cosmic variance, while different particle masses and
spatial resolutions (by a factor of 8 and 2, respectively) of
$\LCDM_{60}$ and $\LCDM_{30}^{ART}$ allow us to check for the
resolution effects.  Comparison of $\xi_{hh}(r,z)$ in the $30\mpch$
runs simulated with different codes and with significantly different
resolutions shows that the two runs agree very well. The
$\LCDM_{30}^{ART}$ simulation had the best mass and force resolution,
yet, we do not find any visible systematic differences at scales
$\lesssim 2\mpch$ with the other two simulations.  There is an
indication that at $r\gtrsim 2\mpch$ the amplitude in the $30\mpch$
simulations is systematically lower than that in the $\LCDM_{60}$,
which can be explained by the effects of finite box size.  This is in
agreement with expectation that the finite size effects become important at
scales $\gtrsim 3\mpch$ ($\approx 0.1$ of the box size).  At smaller
scales, where we expect the $30\mpch$ simulations to produce correct
results, the agreement is very good\footnote{Note that some cosmic
  variance {\em is expected} for these box sizes.}. Besides the
illustration of a very mild evolution of $\xi_{hh}$, Figure 6 also
shows that there is a common feature of the evolution. Although the
exact evolution depends to some extent on the scale, the amplitudes at all
scales are quite high (as high or higher as they are at $z=0$) at very high
redshift ($z\approx 7$).  The amplitude then {\em decreases} until
$z\sim 2-4$, and grows steadily at lower redshifts. It is important to
note that this evolution is more complicated than simple evolution
models often used in the observational and theoretical analyses:
$\xi(r,z)=(r/r_0)^{-\gamma}(1+z)^{-(3-\gamma+\epsilon)}$. Figure 6
shows that parameter $\epsilon$ estimated by such analyses would depend
not only on the redshift range used, but also on the scale at which the
amplitude is measured (as well as on other parameters such as the object's
mass). This calls into question the usefulness of such a simplistic approach
(see also arguments in Moscardini et al.  1998).  Note that there is
also some observational evidence (Giavalisco et al. 1998) indicating
that the above parameterization is a poor description of the observed
galaxy clustering evolution. For a limited range of redshifts $z\sim
0-1$, we find only very weak evolution of halo clustering in {\em
  comoving} coordinates indicating a value of $\epsilon\approx -1$.
This value seems to be favored by observations of galaxy
clustering at these redshifts (e.g., Postman et al. 1998).

{\pspicture(0.5,-1.5)(13.0,17.0)
\rput[tl]{0}(-2.7,17.0){\epsfxsize=14.5cm
\epsffile{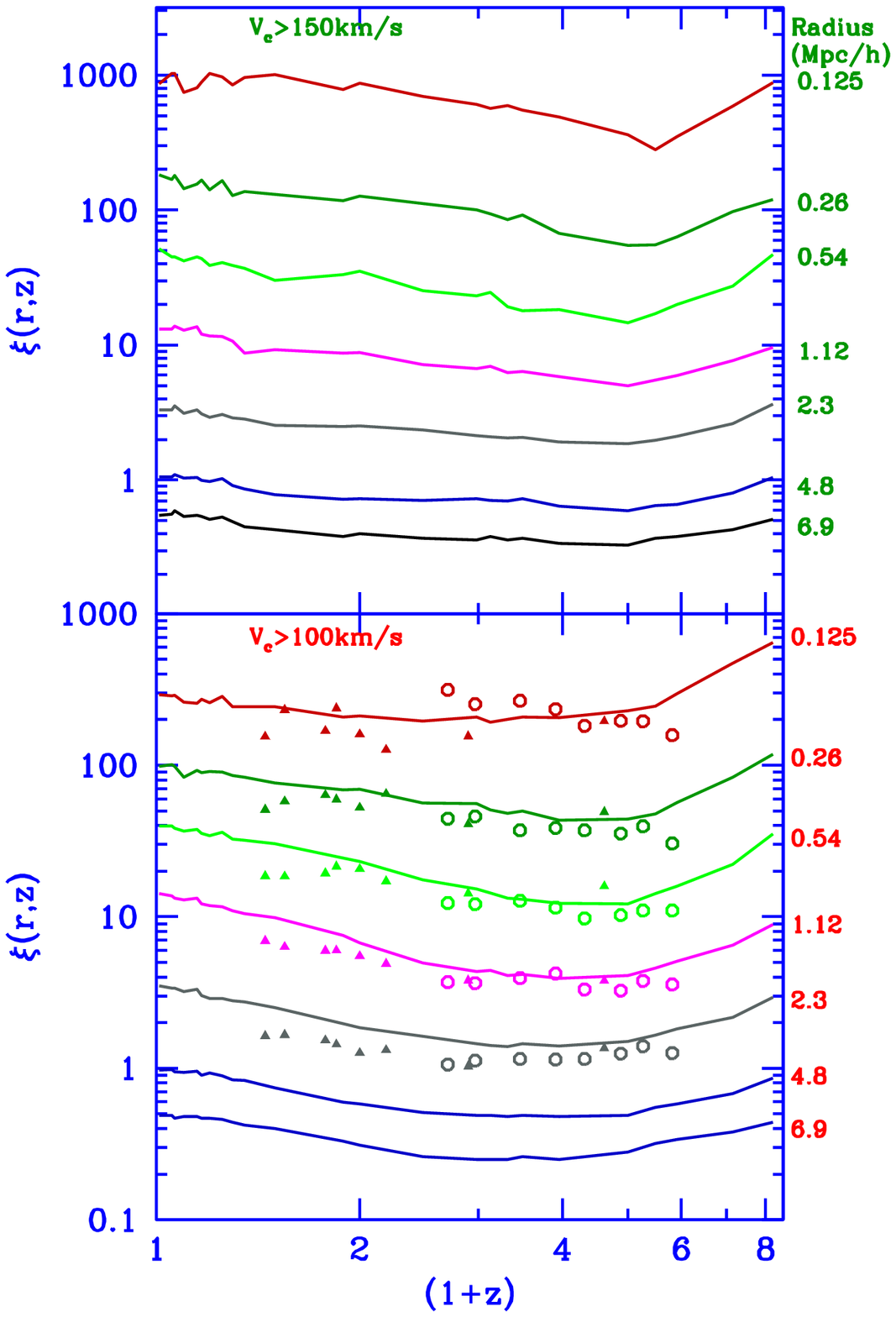}}
\rput[tl]{0}(0.0,2.5){
\begin{minipage}{8.7cm}
  \small\parindent=3.5mm {\sc Fig.}~6.---The evolution of the
  correlation function of halos in \LCDM$_{60}$ simulation at various
  {\em comoving} scales.  Each curve shows the amplitude of the
  correlation function at a fixed comoving radius indicated on the
  right side of the panels. The correlation functions were estimated
  for halos with circular velocity larger than $150\kms$ ({\em top})
  and $>100\kms$ ({\em bottom}). The {\em solid curves} and {\em open
    circles} indicate amplitudes in the ART $60\mpch$ and $30\mpch$
  simulations, respectively; the {\em triangles} show the amplitude in
  the $30\mpch$ {\ap3m} simulation (see \S~5.2 for details and
  discussion).
\end{minipage}
}
\endpspicture}

The power-law fits to $\xi_{hh}$ of the form
$\xi_{hh}(r)=(r/r_0)^{-\gamma}$ for various epochs and for halo
catalogs with different cuts in the maximum circular velocity are
presented in Table 4. Direct (rather than linear fits to
$\lg\xi_{hh}-\lg r$) weighted power-law fits were done to $\xi_{hh}$
with the Levenberg-Marquardt method described in Press et al. (1992).
Each bin of the correlation function was weighted with its Poisson
error $\sigma_{\xi}=(\xi_{hh}+1)/\sqrt{n_p}$, where $n_p$ is the number
of halo pairs in the bin. Visual inspection shows that power-law fits
are very successful at scales $\gtrsim 0.3\mpch$, while at smaller
scales there are $\approx 20-30\%$ deviations in some cases.  The
goodness of the fits is represented in rather small {\em formal} errors
of the best fit parameters $r_0$ and $\gamma$. We have estimated how
these parameters change if the correlation function is re-binned
differently and found that the 
 {\pspicture(0.5,-1.5)(13.0,16.0)
\rput[tl]{0}(-1.0,16.0){\epsfxsize=12cm
\epsffile{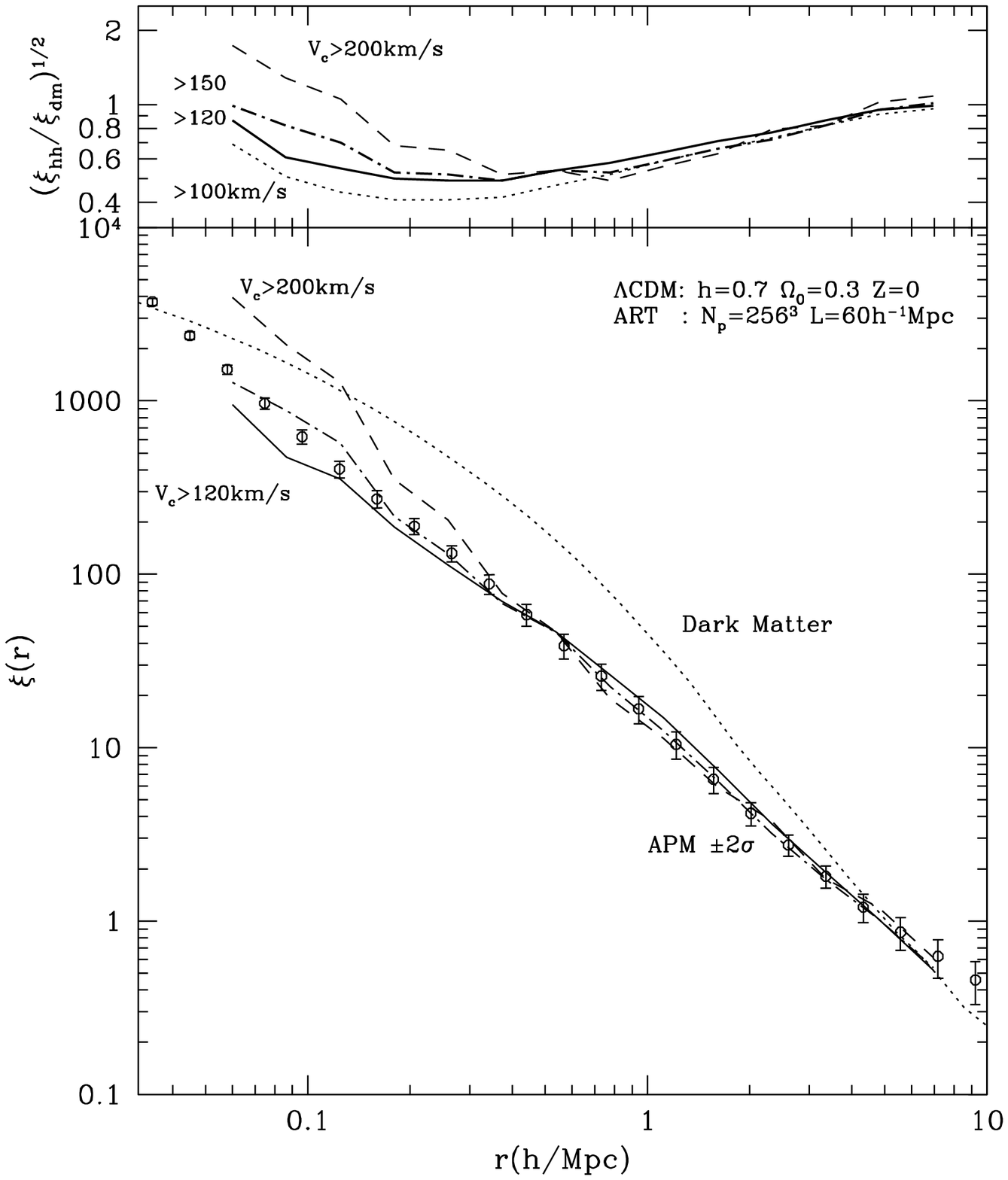}}
\rput[tl]{0}(0.4,3.5){
\begin{minipage}{8.7cm}
  \small\parindent=3.5mm {\sc Fig.}~7.--- {\em Bottom panel:\/}
  Comparison of the halo correlation function in the \LCDM$_{60}$
  simulation with the correlation function of the APM galaxies (Baugh
  1996). Results for halos with maximum circular velocity larger than
  $120\kms$, $150\kms$, and $200\kms$ are presented by the {\em solid},
  {\em dot-dashed}, and {\em dashed} curves, respectively. The {\em
    dotted curve} shows the dark matter correlation function. Note that
  at scales $\gtrsim 0.3\mpch$ the halo correlation function does not
  depend on the limit in the maximum circular velocity (see \S~5.2 for
  details).  {\em Top panel:} Dependence of bias on scale and maximum
  circular velocity. The curve labeling is the same as in the bottom
  panel, except that the {\em dotted line} now represents the bias of
  halos with $V_{max}>100\kms$.
\end{minipage}
 }
\endpspicture}
change is always $\lesssim 3\%$.  The
examination of the Table 4 shows that typical ranges of $r_0$ and
$\gamma$ are $\approx 3-5\mpch$ and $\approx 1.5-1.7$, respectively. For
all halo catalogs, parameters evolve slowly with redshift. The
correlation length $r_0$ decreases somewhat between redshifts of 5 and
3, and then increases steadily until $z=0$. The evolution of $\gamma$
is even slower with the tendency for $\gamma$ to increase by $\approx
10\%$ from $z=5$ to the present epoch. An important and interesting
point is that the correlation amplitude, and hence the value of $r_0$, are
quite different for halo catalogs with different $V_{max}$ cuts. The
correlation lengths for $V_{max}\geq 120\kms$ and $V_{max}\geq 200\kms$
catalogs differ by 25\% at $z\gtrsim 3$, while the difference is only
$\approx 4\%$ at $z=0$. This means there is a mass segregation of
halo clustering properties at high redshift, which, however, is erased
during the subsequent evolution.

During the last few years, there has been tremendous progress in the
observational studies of high-redshift galaxy clustering. We will
discuss how our results on the {\em halo} clustering evolution compare
with the results of observations in \S 6. Here, however, we present a
comparison of the $\xi_{hh}$ with the most accurate measurement of the
galaxy correlation function $\xi_{gg}$ (at $z\approx 0$) made using the
APM galaxy survey (Baugh 1996). Figure 7 shows the $z=0$ correlation
functions of halos and dark matter in the \LCDM\ model and the real
space APM galaxy correlation function. The halo-halo correlation
function was estimated for halo catalogs with cuts in the maximum
circular velocity of $120\kms$, $150\kms$, and $200\kms$. The figure
shows striking agreement between the halo and galaxy correlation
functions: at scales $\gtrsim 0.3\mpch$ the correlation functions of all
halo catalogs match {\em both the shape and the amplitude} of the
$\xi_{gg}$. The correlation function for $V_{max}>150\kms$ catalog
agrees with APM $\xi_{gg}$ within errors at all probed scales. As we
noted above, the differences that exist between the catalogs at high
redshifts virtually vanish during the course of the evolution. This is
manifested in the similarity of $\xi_{hh}$ for different catalogs on
scales $\gtrsim 0.3\mpch$. Note, however, that this does not mean that
$\xi_{hh}$ has no mass dependency. Rather, the result means that by
$z=0$ any mass dependence of the correlation function vanishes when
averaged over a range of galactic masses.  As was explained above, the
poisson errors of the halo correlation functions shown in Figure 7 are
very small and were not shown for clarity. The robustness of the result
can be estimated, however, by comparing $\xi_{hh}$ of $V_{max}>120\kms$
and $V_{max}>150\kms$ catalogs. The number of halos in these catalogs
is significantly different: 4708 and 2480, respectively. This makes the
halo samples largely independent. The correlation functions agree,
however, within the poisson errors.

While the correlation function of halos matches that of galaxies very
accurately, the correlation function of matter $\xi_{dm}$ matches
$\xi_{gg}$ neither in shape nor in amplitude\footnote{This result is
  in agreement with conclusions of Jenkins et al. (1998). As
  was shown, in the previous section, our \xidm\ agrees very well with
  that calculated by Jenkins et al.}. The amplitude is matched only at
scales $\gtrsim 4-5\mpch$. At smaller scales it is much higher than the
amplitude of the APM $\xi_{gg}$, implying that DM halos are {\em
  anti-biased} at these scales with respect to the dark matter.
Moreover, the difference in shape between $\xi_{hh}$ and $\xi_{dm}$
implies that the {\em bias is scale-dependent}.  The scale dependence
of the bias ($\sqrt{\xi_{hh}(r)/\xi_{dm}(r)}$) for the halo correlation
functions is shown in the top panel of figure 7. The bias varies
significantly at scales $\lesssim 5\mpch$ in the range $\sim 0.5-1$.
Moreover, as was shown in Figure 5, the shape of the correlation function 
of dark matter differs from that of $\xi_{hh}$ {\em at all epochs} and
evolves much more strongly than the correlation function of halos. The
former fact implies that {\em the bias is scale-dependent at all
  epochs}, while the latter means that {\em the bias evolves rapidly with
  cosmic time}. The evolution of bias in the $\LCDM_{60}$ run is
illustrated in Figure 8. The evolution is shown for different halo
catalogs and at different scales. The bias evolves very rapidly from
value of $\sim 3-5$ at $z\approx 5$ to $\sim 0.5-1$ at $z=0$. The
evolution depends on the velocity (or mass) cut of the catalog at
high $z$: the halos in the catalogs with higher velocity cuts exhibit 
stronger clustering. This difference vanishes, however, at $z\lesssim
0.5$. The evolution of the scale dependence 
 {\pspicture(0.5,-1.5)(13.0,14.5)
\rput[tl]{0}(-2.2,14.8){\epsfxsize=14cm
\epsffile{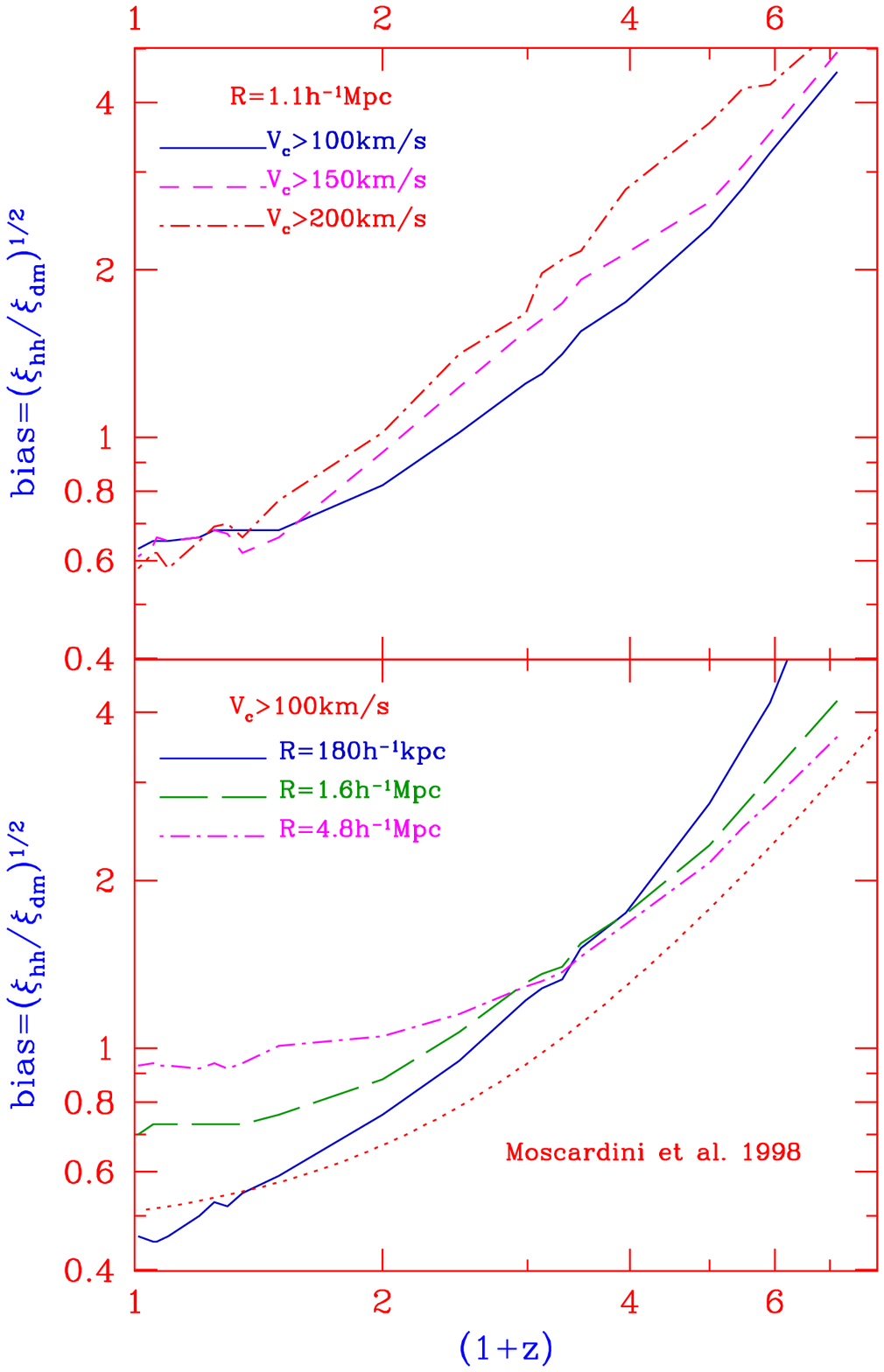}}
\rput[tl]{0}(0.4,0.7){
\begin{minipage}{8.7cm}
  \small\parindent=3.5mm {\sc Fig.}~8.--- {\em Top panel:\/} The
  evolution of bias at {\em comoving} scale of $0.54\mpch$ for halos
  with different lower limit on the maximum circular velocity in the
  \LCDM$_{60}$ simulation. The {\em bottom panel} shows dependence of
  the bias on ({\em comoving}) scale for halos with maximum circular
  velocity $>100\kms$.
\end{minipage}
 }
\endpspicture}
of the bias is also interesting.
At high redshifts the bias was larger at small scales. At small
redshifts the halos are almost unbiased ($b\approx 1$) on a few
megaparsec scales and are anti-biased ($b\approx 0.5-0.6$) on small
scales. 

\begin{figure*}[ht]
\pspicture(0,8.0)(15.0,22.5)
%\psgrid(0,8)(18.8,20)
\rput[tl]{0}(-0.2,22.3){\epsfysize=11cm
\epsffile{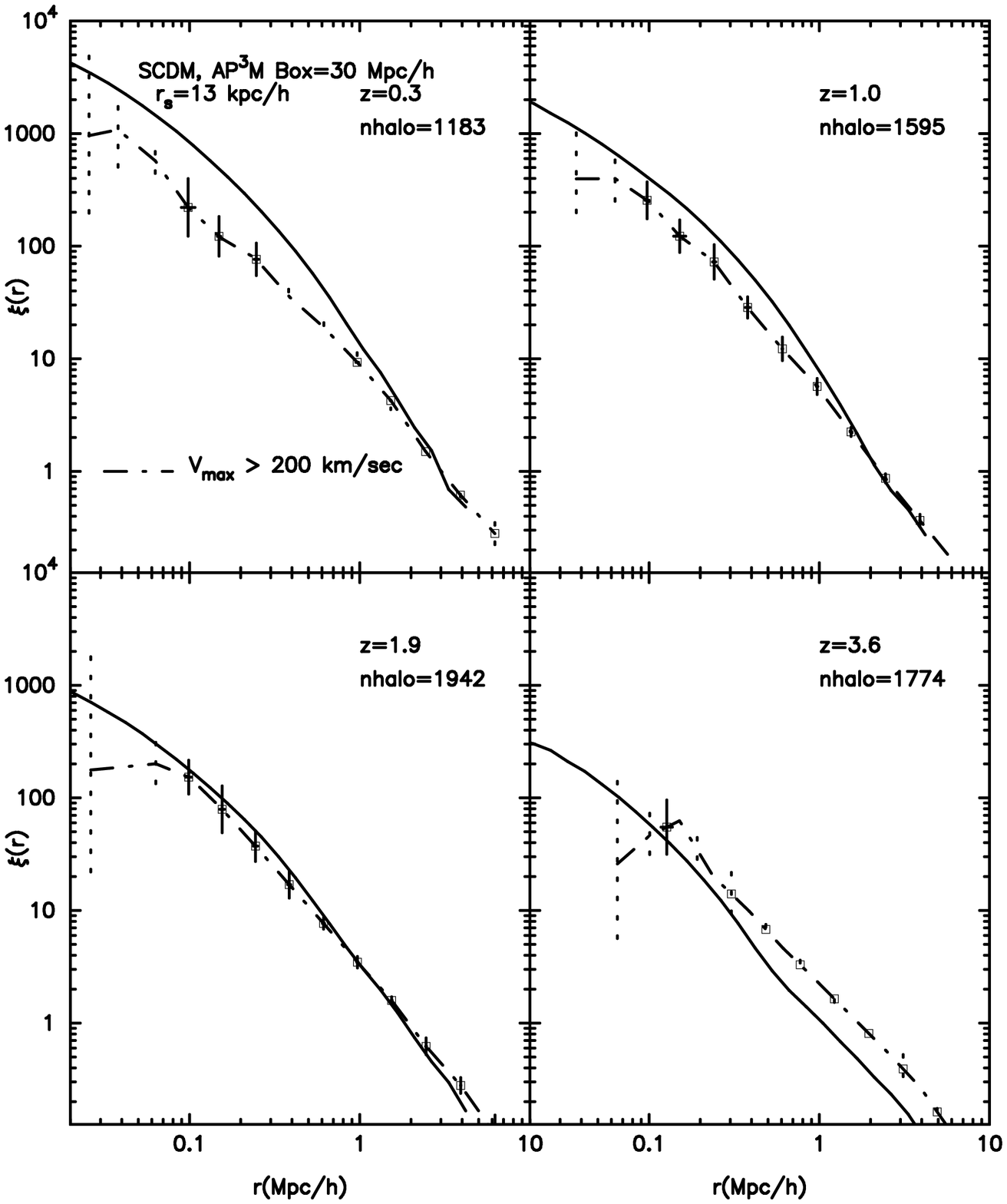}}
\rput[tl]{0}(9.5,22.3){\epsfysize=11cm
\epsffile{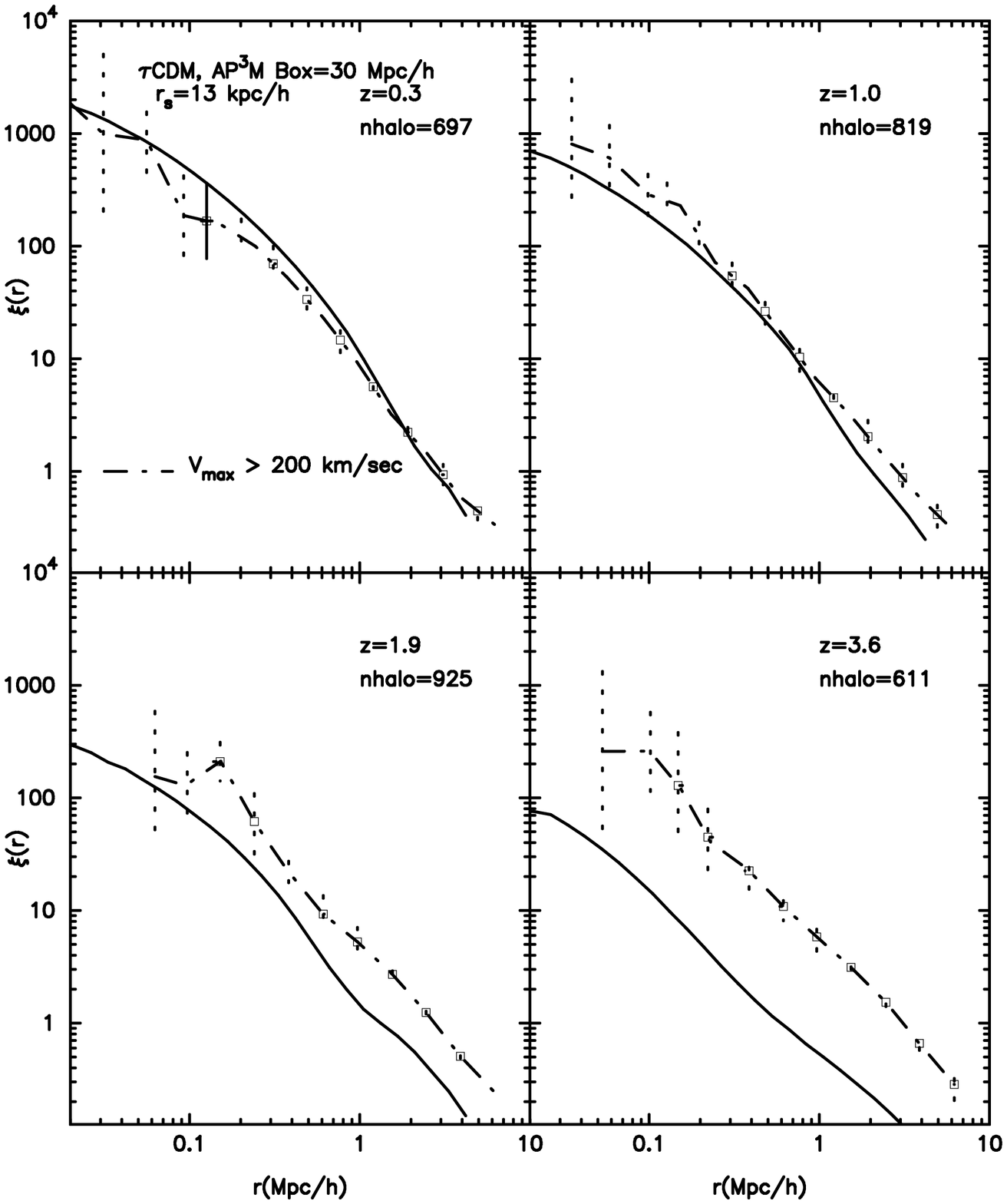}}
\rput[tl]{0}(-0.2,11.0){
\begin{minipage}{8.9cm}
  \small\parindent=3.5mm {\sc Fig.}~9.---The evolution of the
  correlation function of the dark matter ({\em solid lines}) and halos
  ({\em dot-dashed line}) for the SCDM model in the $30\mpch$ \ap3m
  simulation.  The panels show the correlation functions at different
  redshifts. Only halos with maximum circular velocity larger than
  $200\kms$ were used to compute the halo correlation function. The
  number of halos used to estimate the correlation function is
  indicated in each panel.  The poisson errors ({\em dotted}) and
  bootstrap errors ({\em solid}) are shown by vertical bars (see \S~5.3
  for details).
\end{minipage}
}
\rput[tl]{0}(9.6,11.0){
\begin{minipage}{8.9cm}
  \small\parindent=3.5mm {\sc Fig.}~10.--- The same as Figure 9, but
  for the \tCDM model. As in Figure 9, only halos with $V_{max} > 200
  \kms$ were used.
\end{minipage}
}
\endpspicture
\end{figure*}

\begin{figure*}[ht]
\pspicture(0,8.0)(15.0,20.0)
%\psgrid(0,8)(18.8,20)
\rput[tl]{0}(-0.2,19.8){\epsfysize=11cm
\epsffile{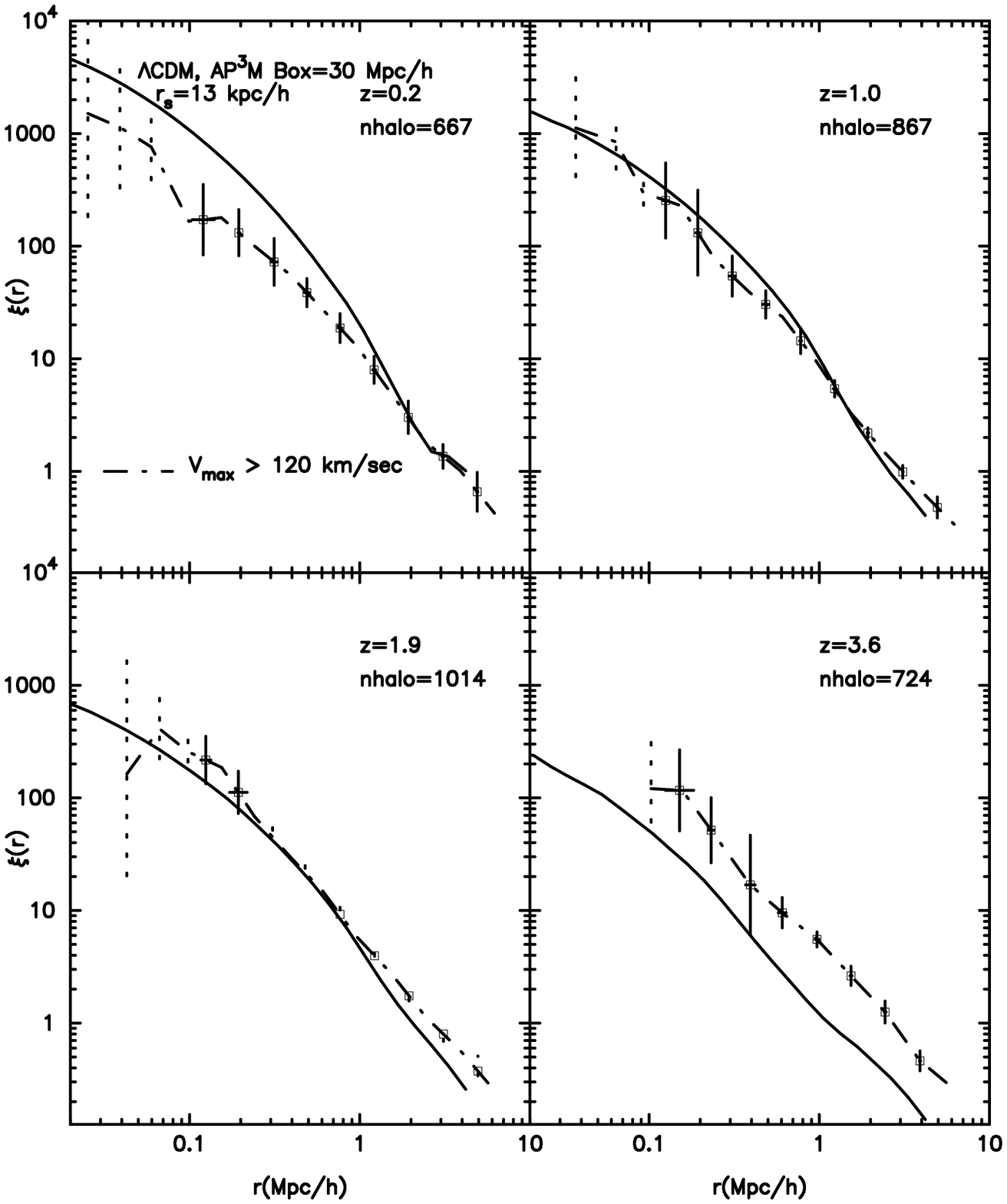}}
\rput[tl]{0}(9.5,19.8){\epsfysize=11cm
\epsffile{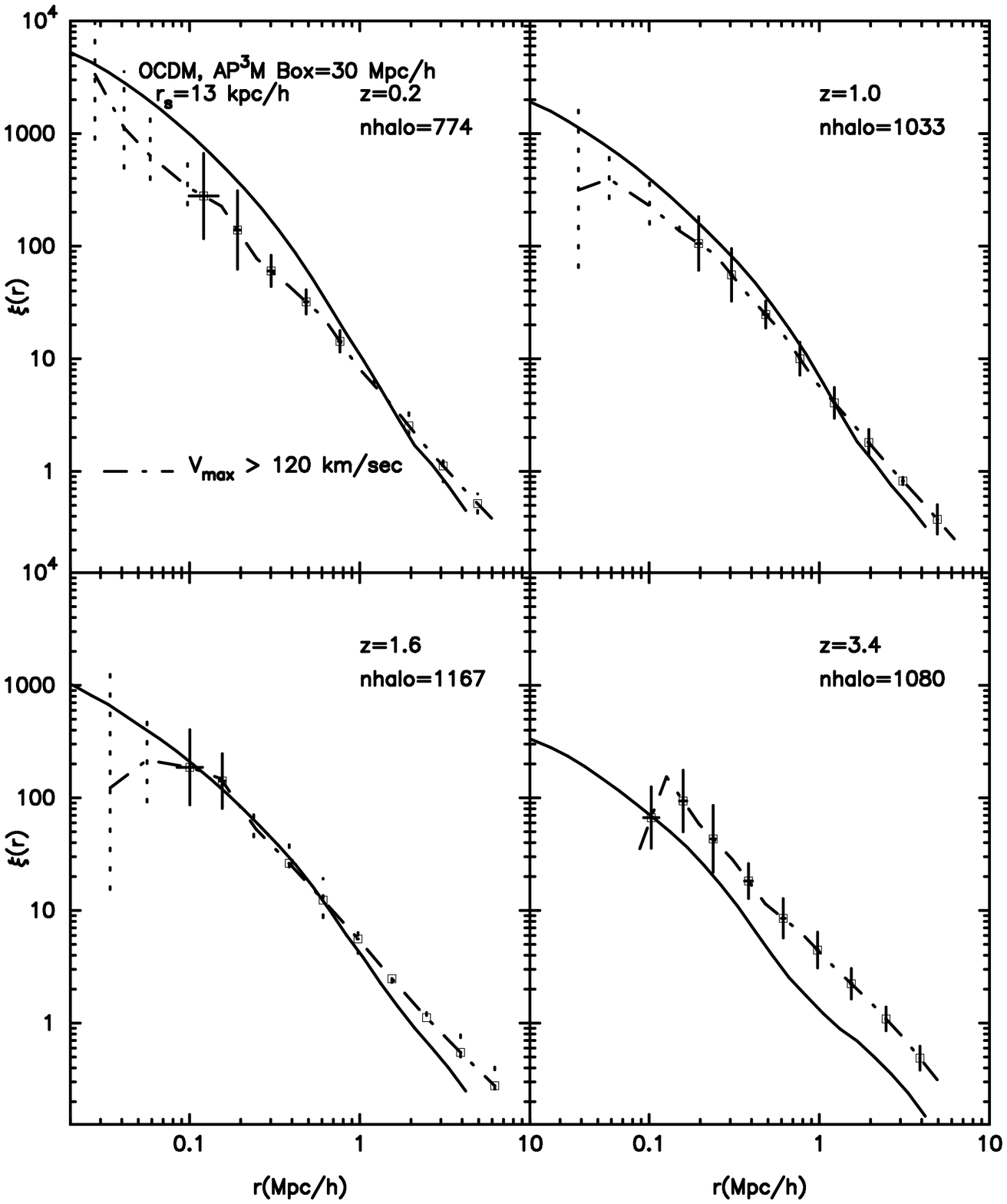}}
\rput[tl]{0}(-0.2,8.5){
\begin{minipage}{8.9cm}
  \small\parindent=3.5mm {\sc Fig.}~11.---The same as Figure 9, but
for the {\LCDM} model. Only halos with $V_{max} > 120 \kms$ were used to
compute the correlation function (see \S~5.3).
\end{minipage}
}
\rput[tl]{0}(9.6,8.5){
\begin{minipage}{8.9cm}
  \small\parindent=3.5mm {\sc Fig.}~12.--- The same as Figure 9, but
for the OCDM model. As in Figure 11, only halos with $V_{max} > 120 \kms$
were used (see \S~5.3).
\end{minipage}
}
\endpspicture
\end{figure*}

\begin{figure*}[ht]
\pspicture(0,8.0)(15.0,24.0)
%\psgrid(0,8)(18.8,20)
\rput[tl]{0}(-1.5,24.0){\epsfysize=12.5cm
\epsffile{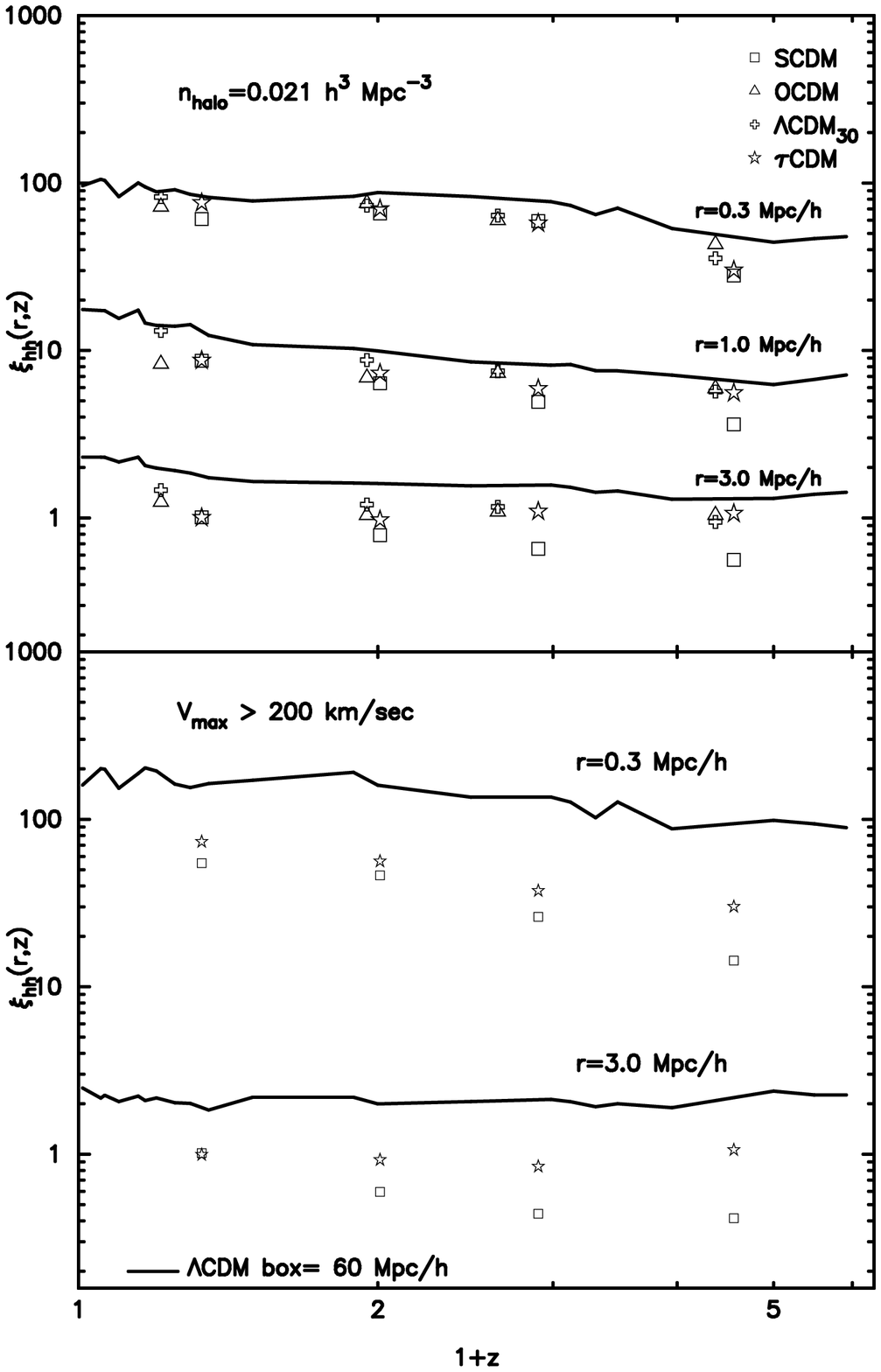}}
\rput[tl]{0}(8.5,24.0){\epsfysize=12.5cm
\epsffile{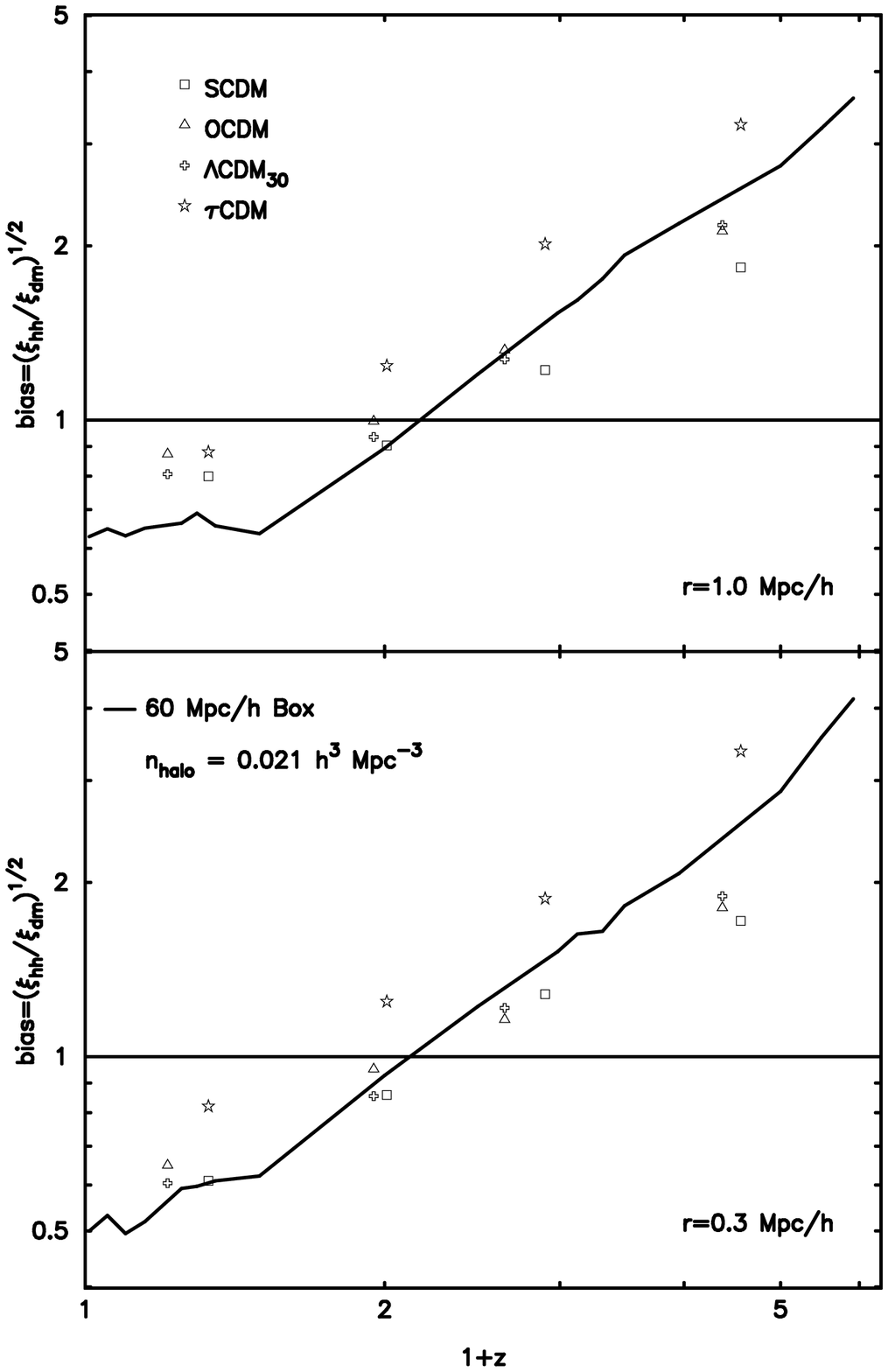}}
\rput[tl]{0}(-0.2,11.2){
\begin{minipage}{8.9cm}
  \small\parindent=3.5mm {\sc Fig.}~13.---The evolution of the halo
  correlation function at various scales for all models.  The
  correlation functions shown in the upper panel were computed using a
  fixed number density of halos (implying different limits on the
  maximum circular velocity cut).  The correlation functions shown in
  the lower panel were computed using all halos with $V_{max}$ larger
  than $200\kms$. Results for only two 30 \mpch\ simulations are
  presented in this panel.  The other two simulations had too small
  numbers of halos.  The solid lines in both panels represent results
  for the \LCDM model with the ART code in the $60\mpch$ box.
\end{minipage}
}
\rput[tl]{0}(9.6,11.2){
\begin{minipage}{8.9cm}
  \small\parindent=3.5mm {\sc Fig.}~14.--- The evolution of bias 
for different models at two scales, $r = 1.0 \mpch$ (upper panel) and
$r= 0.3\mpch$ (lower panel). The markers show results with the \ap3m
code, the solid curves represent results for the \LCDM\ model with the
ART code in the $60\mpch$ box.
\end{minipage}
}
\endpspicture
\end{figure*}

%-------------------------------------------------------------------
\subsection{Evolution of the correlation function and bias in different
  cosmological models}
%-------------------------------------------------------------------

Is evolution of bias observed in the {\LCDM} simulations specific to
this model or is this evolution similar for all of the models? We
address this question by comparing results presented in the previous
section with results of the $30\mpch$ simulations of other cosmological
models (see Table 2).  Figures $9-12$ show the correlation functions of
halos \xihh\ and the dark matter \xidm\ for the four \ap3m $30 \mpch$
runs at four epochs.  Halos with $V_{max} > 200 \kms$ were used to
compute $\xihh(r,z)$ for the SCDM and the \tCDM\ models. A lower
$V_{max}>120\kms$ limit was used for the OCDM and \lcdmtt models. 
The difference
in the $V_{max}$ limits is explained by the difference in the matter
density \ome\ that results in different mass resolution of the
simulations.  In the OCDM and the \lcdmtt runs halos with $V_{max}>200
\kms$ are scarce, while the poorer mass resolution does not allow us to
reduce the limit to 120\kms\ in the SCDM or the \tCDM\ runs.  The total
number of halos found in the simulations is indicated in each panel.
The number depends on the epoch and model and varies from the maximum
of 1942 (in the SCDM run at $z = 1.9$) to the minimum of 611 (in the
\tCDM\ run at $z = 3.6$). The statistics of halos are poorer than in the
{\LCDM$_{60}$} simulation. Therefore, we plot the errorbars associated
with each point of \xihh. We estimate both poisson and bootstrap
errorbars and plot the largest of the two.  The bootstrap error bars
have been estimated as follows. For each run and each epoch we have
drawn 5 randomly selected samples of halos from the corresponding halo
population. The number of halos in each sample is one half of the
total.  We then compute the rms fluctuation between the samples and
divide it by $\sqrt{2}$ to get the 1$\sigma$ errorbar.

Figures $9-12$ show that in all models the evolution of the correlation
function is qualitatively similar to that observed in the \LCDM\
model.
For example, the shape of the \xihh\ is similar (power-law) in all
models and is always different from the corresponding \xidm. This means
that {\em the scale-dependent bias is universal in cosmological
  models}. Details of the evolution are, however, model dependent. The
most drastic differences are seen at the highest redshifts. The
figures, for example, clearly show that the bias at $z=3.4$ has very
different values in different models. While bias in the two low-\ome\ 
models is very similar, the distribution of halos at this redshift is
only weakly biased in the SCDM model, as opposed to the strongly biased
distribution in the \tCDM\ model.  The evolution of the halo
correlation function at three different comoving scales ($0.3$, 1, and
$3\mpch$) for all models is plotted in Figure 13. The solid lines in
both panels represent results for the \LCDM\ model with the ART code in
the $60\mpch$ box. In the upper panel, the evolution is shown for the
halo catalogs with a fixed number density of halos, which was
achieved by varying the $V_{max}$ limit in different models.  Note that
in this case we compare correlations of halo samples with different
mass functions: the \LCDM\ and OCDM halo samples contain many low-mass
halos, while samples in the $\ome = 1$ models contain only massive
halos. Such comparison is interesting for comparisons with observations
when we know the number density of objects in the sample rather than
their mass (or type).  At scales $\lesssim 1\mpch$, differences between
the models are not significant. At 3\mpch\ and at high $z$ the
amplitude of \xihh\ in the SCDM model is significantly lower than in
other models. The amplitude in the rest of the models is surprisingly
similar. Therefore, if the biased galaxy formation scenario is correct
and galaxies can be associated with host halos, this result may have 
interesting implications for the interpretation of clustering observations.
To be able to differentiate between the models, we {\em must} know what
type of the objects was used to estimate the clustering signal. The
knowledge of the number density of objects in the sample is not
sufficient. The point is to some extent illustrated by the lower panel
of Figure 13, where we compare the evolution of the \xihh\ amplitude in
$\LCDM_{60}$, SCDM, and \tCDM\ models for the halo catalogs with the
same selection criterion ($V_{max}>200\kms$).  It is obvious that in
this case the differences between the models are significant. Although
the differences are smaller at low redshifts, at $z\approx 3.5$ the
difference in the amplitude between \LCDM\ and SCDM models is almost an
order of magnitude. This difference can probably be explained by the
delayed formation of galaxy-size halos in the \LCDM\ model as
compared with the SCDM model. The halos in the \LCDM\ form at lower
redshifts with high {\em statistical bias}, while halos in the SCDM
form systematically earlier and thus have had time to go through merging
evolution. The effect of the latter is to decrease the bias (e.g.,
Moscardini et al. 1998). Note also that merging rates are higher in
$\ome=1$ models (e.g., \cite{carlberg90}). These results show that
predictions of cosmological models are very different for samples of
objects selected with the same set of criteria for all models.

The evolution of bias at scales $0.3$ and $1\mpch$ is shown in Figure
14 for all models. Here again we compute the halo correlation function
for the fixed number density of halos. Evolution of bias in all models
is qualitatively similar to that of the \LCDM\ model discussed above:
the bias is a very strong function of redshift.  However, unlike the
\xihh\ amplitude, the value of bias at these scales is very different
among the models. This is not very surprising because when the number
density of halos is fixed, different models have very similar amplitudes
of \xihh\ (see Fig.13), but very different amplitudes of \xidm. The
latter is explained by the differences in the cosmological parameters,
normalization, and the shape of the power spectrum. A more interesting
implication of the Figure 14 is that differences in bias get smaller at
low redshifts, virtually disappearing at $z=0$. The same effect can be
observed in the evolution of the amplitude in Figure 13. As we will
argue in the next section, the evolution of the halo correlations and
bias at these scales is likely to be driven by the halo dynamics within
nonlinear structures, in which case the differences between different
cosmologies are largely erased. The evolution shown in Figures 13 and
14 provides, therefore, indirect support for this point: the
differences in clustering amplitude and bias between the models
disappear at $z\lesssim 1$, where most of the clustering signal comes
from the halos located in nonlinear structures.

%===================

\section{Discussion}

%===================

It is interesting to compare the evolution of the halo correlation function
and the bias observed in our simulations with predictions of the analytical
models and results of previous numerical simulations.  The fact that
clustering strength of halos at high redshifts is comparable to that at
the present epoch, has been noted in results of many simulations (e.g.,
\cite{davis85}; Brainerd \& Villumsen 1994; Col\'in et al. 1997; and
references therein). Bagla (1998) summarizes the generic behaviour of
the correlation amplitude of halos above certain mass: the amplitude is
high at very high redshifts, when halos are being formed, decreases
thereafter and reaches a minimum, and then increases slowly and
steadily until the present epoch. The results presented in the previous
section (see Figs. 6 and 13) are in agreement with this picture. Thus,
there seems to be a good qualitative (although, in some cases, not
quantitative) agreement among results of different numerical simulations
concerning the evolution of the halo correlation function.

Analytical models have reached a sufficient degree
of sophistication to be able to predict the evolution of halo clustering
in mildly nonlinear regimes (see \S~2).  The halos are found to form at
the peaks of the density field (e.g., \cite{frenk88}) and their bias
exhibits a simple scaling relation with the height of these peaks
(Kaiser 1984; \cite{bbks86}).  At any given epoch the halo population
represents a mix of halos formed at different redshifts: newly-born or
already evolved through merging.  The evolution of the correlation 
of halos in such a hierarchical framework is described using the extended
Press-Schechter formalism (MW). To compare predictions of
the analytical models with our results, we will use the approximation to
the evolution of the {\em effective bias} given by eq. (4) (Moscardini et al.
1998). This approximation describes the evolution of bias of a sample of
all halos above a certain mass. This is roughly equivalent to our
definition of halo samples with a limiting maximum circular velocity.
The prediction of this approximation is shown in Figure 8 with the
dotted line, where we used $b_{eff}(0)=0.51$ and $\beta=1.90$ (see
eq.[4]) appropriate for our {\LCDM} model and for the mass limit of
$M\geq 10^{11}h^{-1} {\rm M_{\odot}}$ (Moscardini et al. 1998). The
analytical model is expected to provide a good approximation at scales
where $\xi_{hh}\lesssim 1$ (Mo et al. 1996), i.e. at $r\gtrsim
4-5\kpch$ (see Table 4). Therefore, the analytical prediction should be
compared with the curve showing evolution of bias at $r=4.8\mpch$.

The comparison shows that both the numerical result and the analytical
model predict a rapid decrease of bias with cosmic time.  Moreover, at
high redshifts ($z\gtrsim 3$) they agree well {\em quantitatively}.  At
lower redshifts, however, the two predictions deviate from each other
and are different by a factor of 2 at $z=0$: at $r=4.8\mpch$ almost no
bias is observed in the simulation, while strong anti-bias of $b\approx
0.5$ is predicted by the analytical model.  The differences are not
surprising, given the differences in our definition of a halo from that
of the Press-Schechter halo. The definition of the latter does not
include ``satellite'' halos; a halo ceases to exist, once it becomes
bound to another halo (i.e., {\em ``merges''}) and orbits inside that
halo's virial radius. Our definition, on the other hand, does take
satellite halos into consideration, because we include in our halo list
every gravitationally bound clump of particles, regardless of whether
it is also bound to a larger system or not. The similarity between our
numerical result and the prediction of the analytical model at $z\gtrsim 3$
is then an indication that the two definitions are equivalent at these
high redshifts. Indeed, large systems such as clusters and groups have
not yet formed at these redshifts, and the fraction of satellite halos
in our catalogs (i.e., all halos above a certain mass limit) is
relatively small.  At smaller redshifts the ever larger fraction of
halos become satellites to more massive halos and the two halo
definitions result in rather different halo samples. This explains the
large difference predicted for the value of bias at $z=0$. We have
found that in our simulation the $z=0$ amplitude of the 2-point
correlation function of the Press-Schecter halos (i.e., {\em isolated}
in terms of their virial radius) at $r=4.8\mpch$ is approximately {\em
  two times as small} as the amplitude of the correlation function of the
BDM halos shown in Fig. 7, resulting thus in the {\em twice as small
  bias} of $b\approx 0.5$. In this respect, we believe that there is no
contradiction between our results and predictions of the analytical
model, once the difference in the halo definition is taken into
account. This also explains the difference in the bias value with the
result of Jing (1998), who found bias of $b\approx 0.5-0.7$ for the
galaxy-size halos defined with the FOF algorithm\footnote{In this case
  again the halos are identified as isolated density peaks with
  an overdensity exceeding the value expected of the virialized object. No
  satellite halos can be identified with this definition.}. We find
also that, contrary to the assumption of the analytical models, the
{\em small-scale} bias of the halo distribution is {\em
  scale-dependent} regardless of the halo definition. The large-scale
bias ($r\gtrsim 5\mpch$), on the other hand, is not probed in our
simulations and may be independent of scale, in accord with assumptions
of the analytical models. This indeed was suggested by the simulations
of Jing (1998). We think that the most encouraging result is the
agreement between numerical and analytical modeling on the general form
of the bias evolution demonstrated in Fig.~8. This indicates that we now
have a solid general understanding of the nature of bias and of the
processes driving its evolution at redshifts $z\gtrsim 2-3$.

At smaller redshifts, the merging rate is considerably smaller and
small-scale correlations are sensitive to the dynamics of halos
inside the nonlinear structures. Particularly, the dynamics and the clustering
evolution of satellite halos in high-density regions are essentially
independent of the background cosmology and are driven by such processes
as dynamical friction and tidal stripping (e.g., KGKK).
These processes tend to suppress the growth of the
correlation amplitude, thus counteracting the clustering growth due to
the gravitational pull. This leads to the anti-bias observed in our
simulations at nonlinear scales and at small redshifts. Indeed, the
correlation amplitude at small scales ($r\lesssim 5\mpch$) is
approximately constant at redshifts $z\approx 0-1$ (see Figs. 6 and 13).
Moreover, anti-bias is observed in all cosmological models studied
in this paper (see Figs. 9-12).  The fact that differences in the correlation
amplitude and bias, existing between the cosmological models at
$z\approx 3$, virtually vanish by the present epoch (see Figs. 13 and 14)
argues for the importance of the nonlinear halo dynamics. This result
also implies that low-redshift clustering depends only weakly on the
background cosmology. Therefore, the information about the underlying
cosmological model can probably be extracted only from the high-$z$
($z\gtrsim 3$) clustering data.  As was discussed in \S~4.3, the
numerical resolution required to assure the survival of halos in 
high-density regions is high, and has not been reached in previous
simulations. Our results therefore indicate that high resolution is
important for correct modelling of the bias evolution at small
redshifts.

Although our definition of a halo is different from that used in the
conventional Press-Schechter framework, we believe that it is closer to
what can be identified in a simulation as a galaxy location. It seems
likely that in {\em every} sufficiently massive ($M\gtrsim
10^{11}h^{-1} {\rm M_{\odot}}$) {\em gravitationally bound} halo
baryons will cool, form stars, and produce an object ressembling a
galaxy (e.g., Kauffman, Nusser \& Steinmetz 1997; Roukema et
al. 1997; Yepes et al.  1997; \cite{sp97}). We believe, therefore, that
each of the halos in our catalog can be associated with a
``galaxy''\footnote{We cannot, of course, unambiguously  assign type,
  color, luminosity, or other galactic properties to our halos without
  additional modelling. Our results, therefore, should be applicable to
  ``global'' galaxy surveys, such as the APM survey, in which galaxies
  are selected solely on the basis of their luminosity (loosely related
  to the mass).}. Observationally, many distinct galaxies are located
well inside the virial radii of massive galaxies, groups, and clusters.
These galaxies, non-existent by definition in the ``virial
overdensity'' halo catalogs, are included in galaxy surveys and are
used to compute the correlation function. Our definition, therefore, is
natural, if the goal is to compare the observations with predictions of
the numerical simulations. Although, as we discussed above 
(\S~4.3), the resolution (and, correspondingly, the computational costs)
required for galaxy-size halos to survive in the tidal fields of
high-density regions is quite high, the subsequent comparison with the
observations is straightforward and does not require ambiguous
corrections for the ``overmerging''.

In \S 5.2 we have compared the correlation function of halos $\xihh$ in
our $60\mpch$ {\LCDM} simulation with the $z=0$ correlation function of
galaxies $\xigg$ (Baugh 1996) in the APM catalog (Loveday et al. 1995)
and found a very good agreement between the two. The APM galaxy
correlation function is measured very accurately, which makes the
agreement within the $1\sigma$-errorbars very striking (see Fig.7).
Recently, the advent of new faint galaxy surveys allowed the
measurement of the clustering surveys at redshifts of $z\approx 0-1$
(e.g., Le F\`evre et al. 1996; Shepherd et al. 1997; Carlberg et al.
1997; Connolly et al.  1998; Postman et al. 1998; Carlberg et al.
1998). Unfortunately, there is some disagreement between these studies
concerning the amplitude of the clustering signal, which possibly indicates
that there is morphology and/or luminosity segregation in
the clustering of the intermediate redshift galaxies. It is indeed
true that the correlation amplitude depends on the luminosity of the
galaxies (e.g., Carlberg et al. 1998; Postman et al. 1998).  The
comoving correlation length $r_0$ measured in these surveys varies,
generally lying in the range $r_0\sim 2-4\mpch$ with an average value of
about $r_0\approx 3\mpch$. The correlation length of bright galaxies
is, however, somewhat larger and consistent with the correlation
length of the local galaxies $r_0\approx 5\mpch$ (Carlberg et al. 1998;
Postman et al. 1998).  Comparison of the above values of $r_0$ with
correlations in our \LCDM\ simulation (Table 4), shows that we predict
a correlation length in good agreement with that of bright galaxies and
somewhat larger than the $r_0$ value of the faint galaxies.  The
latter, however, is measured with rather large uncertainties and our
values of $r_0$ are actually consistent with the observed ones within
$1\sigma$.

Remarkable progress of the high-redshift galaxy detection techniques,
based on the search for the signatures of the Lyman break in the colors
of faint galaxies (Steidel \& Hamilton 1992, 1993), resulted in a rapid
growth of amount and quality of the clustering observations at
$z\approx 3$ (Steidel et al. 1998; Giavalisco et al. 1998; Adelberger
et al. 1998). These ``Lyman break galaxies'' (LBGs) were found to be
clustered at $z\approx 3$ as strongly as the present day galaxies. The
real-space correlation function of these galaxies was well-described by
the conventional power-law form with the value of the slope $\gamma$
and the correlation length $r_0$ consistent with the $z\approx 0$
values (Giavalisco et al. 1998): $\gamma\approx 1.7-2.2$ and
$r_0\approx 3\mpch$. These values are in reasonably good agreement with
$z=3$ values for our $V_{max}>120 {\rm km/s}$ catalog (Table 4).  The
value of bias that is measured for the LBGs at scales of $\approx
5-10\mpch$ is $b\sim 1.5, 3.6, 4.5$, for the $\ome=1$, $\ome=0.2$
(open), and $\ome=0.3$ (flat) models, respectively (Giavalisco et al.
1998; Adelberger et al. 1998). These values can be compared with
$b(z=3)$ for halos in our simulations, shown in Figs.~8 and 14.  All
models agree with the observations, within the uncertainties of the
galaxy-to-halo mapping.  This result is in general agreement with the
results of the numerous recent numerical studies that modeled the
clustering of LBGs (e.g., Wechsler et al.  1998; Jing \& Suto 1998;
Bagla 1998; Governato et al. 1998). However, there appear to be some
puzzling details in comparisons with the data. The observed value of
$b_{LBG}\approx 3-4$ can be reproduced in the {\LCDM} for massive halos
with $V_{max}>200 {\rm km/s}$ (see top panel of Fig.8). This is in
agreement with almost all other theoretical studies. However, the
correlation function of the LBGs measured by Giavalisco et al. (1998),
does not agree with the correlation function measured for the
$V_{max}>200 {\rm km/s}$ of our {\LCDM} halos: $r_0\approx 3\mpch$ for
LBGs versus $r_0\approx 4.5\mpch$ for the halos. This disagreement was
actually noticed in the study of Adelberger et al. (1998) who used the
count-in-cells analysis to derive the value of bias. The values of the
parameters of the correlation function that were derived from the
observed rms fluctuations of galaxies in cells of $\approx 12\mpch$
($\sigma_{gal}\approx 1.1\pm 0.2$), are considerably higher than those
measured directly by Giavalisco et al. (1998). The corresponding value
of $\sigma_{halo}(r=12\mpch)$ in our {\LCDM} simulation is $\approx
0.6, 0.7, 0.9$ for halos with $V_{max}>120, 150, 200 {\rm km/s}$,
respectively. This is consistent with the interpretation of the LBGs as
objects residing inside massive ($V_{max}>200 {\rm km/s}$ or $M\gtrsim
10^{12}h^{-1} {\rm M_{\odot}}$) DM halos. This result supports the
interpretation of Adelberger et al. (1998) and suggests that the
correlation amplitude of the LBGs may be higher than that obtained from
the observed angular correlation function (Giavalisco et al. 1998).

Overall, we believe that the comparisons discussed above indicate that
there is good agreement between our results and the clustering data
at both low and high redshifts. This implies that hierarchical models
in which observed galaxies form in the host DM halos naturally explain 
the observed galaxy clustering at different epochs, including excellent
agreement with the accurately measured $z=0$ correlation function. On
the other hand, the generic form of the bias evolution observed in the
numerical simulations at high redshifts agrees well with the prediction
of the analytical models based on the extended Press-Schechter
formalism. This implies that we understand the nature of the bias and
the processes that drive its evolution at high $z$. At low redshifts,
the bias evolution of gravitationally bound halos is driven by the
dynamical processes inside the nonlinear structures which are largely
independent of cosmology. The study of these processes is important for
a successful modelling of galaxy clustering at $z\lesssim 1$.

%================

\section{Summary}

%================

We have studied the evolution of the correlation function and bias of
galaxy-size halos in different cosmological models (\LCDM, OCDM, \tCDM,
and SCDM). The high-resolution of our numerical simulations allowed us
to avoid the overmerging in the high-density regions and estimate the
correlation amplitude and bias {\em directly} at small (down to $\sim
100\kpch$) scales. The main results and conclusions presented in this
paper are as follows. 

1. At all epochs, the 2-point correlation function of galaxy-size halos
$\xihh$ is well approximated by a power-law $\xihh=(r/r_0)^{-\gamma}$
with the slope $\gamma\approx 1.6-1.8$. The correlation length $r_0$ at
$z=0$ is $\approx 5\mpch$, regardless of the minimal mass limit of the
halo samples. At high redshifts, the correlation function evolves
non-monotonically: $r_0$ decreases somewhat between redshifts of 5 and
3, and then increases steadily until $z=0$. For the most massive halos,
the correlation length at $z\approx 5$ is comparable to that at
$z=0$.

2. The difference between the shape of the $\xihh$ and the shape of the
correlation function of matter results in a {\em scale-dependent
  bias} at scales $\lesssim 7\mpch$. We find this to be a generic
prediction of the hierarchical models independent of the epoch and of
the model details.

3.  Another generic prediction is that the {\em comoving} amplitude of
the correlation function for halos above a certain mass evolves
non-monotonically: it decreases from an initially high value at $z\sim
3-7$, and very slowly increases at $z\lesssim 1$. This behaviour at
large scales was demonstrated by a number of authors (see \S~6). Here,
we have shown that this behaviour also applies to the correlation
amplitude at small scales ($\lesssim 1\mpch$). The non-monotonic
evolution of the correlation function calls into question the usefulness of the
simplistic ``$\epsilon$-models'' as a description of the clustering
evolution. We note, however, that at $z\lesssim 1$ the evolution of the halo
correlation function is approximately monotonic (albeit dependent on
scale). The very slow evolution of the halo correlation amplitude in {\em
  comoving} coordinates at these redshifts implies a value of
$\epsilon\approx -1$, which is in agreement with the values preferred by
the observations (e.g., Postman et al. 1998).

4. The evolution of the halo correlation function is {\em very mild}
compared to evolution of the dark matter correlation function. The
latter evolves by a factor of $\sim 10-60$ (depending on scale) between
redshifts of $\approx 7$ and $0$, while the difference in amplitude of
the former between any two epochs is less than a factor of 2. The large
difference in the evolution rates of the matter and halo correlation
functions means that the bias evolves rapidly with cosmic time: it
changes from high $b$ values of $\sim 2-5$ at $z\sim 3-7$ to {\em
  anti-bias} $b$ values of $\sim 0.5-1$ on small $\lesssim 5\mpch$ scales at
$z=0$.

5. We find that our results agree well with existing clustering data at
different redshifts, indicating a general success of the hierarchical
models of structure formation in which galaxies form inside the host DM
halos.  Particularly, we find excellent agreement in both slope and
amplitude between $\xihh(z=0)$ in our \lcdmxx simulation and the galaxy
correlation function measured using the APM galaxy survey. At high
redshifts, all models reproduce well the observed
clustering of the Lyman-break galaxies.  Our results imply that for
high-redshift clustering to be used as a cosmological test, it is
crucial that we know what type of objects are used to estimate the
clustering signal. The knowledge of the number density of objects {\em
  is not sufficient} (see \S~5).

6. We find good agreement at $z\gtrsim 2$ between our results and
predictions of the analytical models of bias evolution (MW; Matarrese
et al. 1997). This indicates that we now have a solid understanding of
the nature of the bias and of the processes that drive its evolution at
these $z$. We argue, however, that at lower redshifts the evolution of
bias is driven by dynamical processes, i.e., dynamical friction and
tidal stripping, inside the nonlinear high-density regions such as
galaxy clusters and groups.  These processes do not depend on cosmology
and tend to erase the differences in clustering properties of halos
that exist between cosmological models at high $z$. The latter result
implies that low-redshift clustering is probably not a very strong
discriminator between cosmological models. 

We believe that the success of the current theoretical models in
interpreting the clustering data forms a solid foundation for
further sophistication of the models by including the processes
important for galaxy formation (such as dynamics of baryons,
cooling, star formation, and stellar feedback). These models would allow
one to predict the observed properties of galaxies and thus mimick the
observational selection criteria, allowing for a robust comparison
between the model and the data. We believe that the differences between
high-$z$ clustering properties of objects in different cosmological
models demonstrated in this study (see \S~5), the improved theoretical
models, and the ever increasing amount of new clustering data can be
successfully combined in the near future to put useful constraints on
the cosmological parameters describing our Universe.

\acknowledgements This work was funded by NSF grant AST-9319970, NASA
grant NAG-5-3842, and NATO grant CRG 972148 to the NMSU. P.C. was
partially supported by DGAPA/UNAM through project IN-109896.  We are
grateful to Stefan Gottl\"ober for the help in computing the
correlation function of isolated halos, to Carlton Baugh for sending us
the APM correlation function in electronic form, and to Adrian Jenkins
for providing the Virgo dark matter correlation function in electronic
form. We thank Elizabeth Rizza for the help in improving the language
of the manuscript.  Our ART simulations were done at the National
Center for Supercomputing Applications (Urbana-Champaign, Illinois) and
on the Origin2000 computer at the Naval Research Laboratory.  The \ap3m
simulations have been carried out on the Origin-2000 at the Direcci\'on
General de Servicios de C\'omputo, UNAM, Mexico and at the DEC Alpha
Stations at CITA (Toronto, Canada).  P.C. is very grateful to R.
Carlberg for an access to the CITA computing resources, where part of
the \ap3m simulations were run.
%===================================================================

\end{document}